\documentclass[final,comsoc,twocolumn]{IEEEtran}

\IEEEoverridecommandlockouts

\usepackage{graphicx}
\usepackage{color}
\usepackage{cite}
\usepackage{amsmath}
\usepackage[caption=false]{subfig}

\usepackage{amssymb}
\usepackage{amsfonts}

\usepackage{algorithmic}
\usepackage{algorithm}

\usepackage{amsmath}

\usepackage{url}
\newcommand{\subparagraph}{}
\usepackage{amssymb}
\usepackage{amsfonts}
\usepackage{lineno}
\usepackage{comment}

\usepackage{siunitx}

\newcounter{tempEquationCounter}
\newcounter{thisEquationNumber}
\newenvironment{floatEq}
{
	\setcounter{thisEquationNumber}{\value{equation}}\addtocounter{equation}{1}
\begin{figure*}[!t]
\normalsize\setcounter{tempEquationCounter}{\value{equation}}
\setcounter{equation}{\value{thisEquationNumber}}
}
{
\hrulefill\vspace*{4pt}
\end{figure*}
}

\newtheorem{theorem}{\bf Theorem}

\newtheorem{definition}{\bf Definition}

\newcommand{\sue}{c} 
\newcommand{\us}{l} 
\newcommand{\pc}{p}
\newcommand{\op}{\beta} 
\newcommand{\EP}{\zeta} 
\newcommand{\Ns}{\emph{N}_s} 

\def\mathbi#1{\textbf{\em #1}}
\begin{document}
\vspace{-6cm}
\title{Joint Load Balancing and Interference Mitigation in 5G Heterogeneous Networks} \vspace{-5cm}
\author{Trung~Kien~Vu,~\IEEEmembership{Student~Member,~IEEE,}
        Mehdi~Bennis,~\IEEEmembership{Senior~Member,~IEEE,}
        Sumudu~Samarakoon,~\IEEEmembership{Student~Member,~IEEE,}
        M{\'e}rouane~Debbah,~\IEEEmembership{Fellow,~IEEE,}
        and~Matti~Latva-aho,~\IEEEmembership{Senior~Member,~IEEE}

\thanks{Manuscript received Nov 14, 2016; revised April 09, 2017 and May 26, 2017; accepted June 13, 2017; Date of publication June 27, 2017; The authors would like to thank the Finnish Funding Agency for Technology and Innovation (Tekes), Nokia, Huawei, and Anite for project funding. The Academy of Finland funding through the grant 284704 and the Academy of Finland CARMA project are also acknowledged. The research of M. Debbah has been supported by the ERC Starting Grant 305123 MORE (Advanced Mathematical Tools for Complex Network Engineering). This paper was presented in part at the 22th European Wireless Conference, Oulu, Finland, May 2016~\cite{Vu2016}. The associate editor coordinating the review of this paper and approving it for publication was S. Jin (\it{Corresponding author: Trung Kien Vu.})}

\thanks{T. K. Vu, S. Sumudu, and M. Latva-aho are with the Centre for Wireless Communications, University of Oulu, 90014 Oulu, Finland, (email: \{trungkien.vu, sumudu.samarakoon, matti.latva-aho\}@oulu.fi).}

\thanks{M. Bennis is with the Centre for Wireless Communications, University of Oulu, 90014 Oulu, Finland, and also with the Department of Computer Engineering, Kyung Hee University, Yongin 446-701, South Korea (e-mail: mehdi.bennis@oulu.fi).}

\thanks{M. Debbah is with the Large Networks and System Group (LANEAS), CentraleSup\'elec, Universit\'e Paris-Saclay, 91192 Gif-sur-Yvette, France and is with the Mathematical and Algorithmic Sciences Laboratory, Huawei France R\&D, 92100 Paris, France  (e-mail: merouane.debbah@huawei.com).}

\thanks{Citation information: DOI: 10.1109/TWC.2017.2718504, IEEE Transaction on Wireless Communications} }\vspace{-6.0cm}

\maketitle

\begin{abstract}
We study the problem of joint load balancing (user association and user scheduling) and interference management (beamforming design and power allocation) in heterogeneous networks (HetNets) in which massive multiple-input multiple-output (MIMO) macro cell base station (BS) equipped with a large number of antennas, overlaid with wireless self-backhauled small cells (SCs) are assumed. Self-backhauled SC BSs with full-duplex communication employing regular antenna arrays serve both macro users and SC users by using the wireless backhaul from macro BS in the same frequency band. We formulate the joint load balancing and interference mitigation problem as a network utility maximization subject to wireless backhaul constraints. Subsequently, leveraging the framework of stochastic optimization, the problem is decoupled into dynamic scheduling of macro cell users, backhaul provisioning of SCs, and offloading macro cell users to SCs as a function of interference and backhaul links. Via numerical results, we show the performance gains of our proposed framework under the impact of SCs density, number of BS antennas, and transmit power levels at low and high frequency bands. It is shown that our proposed approach achieves a $5.6\times$ gain in terms of cell-edge performance as compared to the closed-access baseline in ultra-dense networks with $350$ SC BSs per $\text{km}^2$.
\end{abstract}

\IEEEpeerreviewmaketitle
\section{Introduction}
\label{Intro}
To meet the massive data traffic demands in next generation $5$G wireless networks a number of emerging technologies are currently investigated: $1$)  higher frequency spectrum (mmWave); $2$) advanced spectral-efficiency techniques (massive MIMO); and $3$) ultra-dense small cell deployments~\cite{Nokia2011}. In this paper, we focus on the interplay between massive MIMO and a dense deployment of SCs in higher frequency bands. Massive MIMO plays an important role in wireless networks due to an improvement in energy and spectral efficiency~\cite{marzt2010non}. In massive MIMO, a macro base station (MBS) equipped with a few hundreds antennas simultaneously serves tens of user equipments (UEs) and provides wireless backhaul to SCs, while the remaining degree of freedom of massive MIMO can be used to mitigate the cross-tier interference. Ultra dense SC deployment provides an effective solution to increase network capacity by a factor of $100\times$ or more and offloads the wireless data from the MBS~\cite{vu2015cooperative}. In order to reduce the deployment cost of SC, wireless backhaul has been considered as an attractive solution. In parallel to that, recent advances in full-duplex (FD) enables doubling spectral efficiency and lowering latency in which FD-enabled SCs relay data from the massive MIMO MBS to the UEs in the same frequency band~\cite{li2015small}.

MmWave with short wavelength enables Massive MIMO to pack more antennas into highly directional footprint and to smartly do beamforming~\cite{hur2013millimeter}, making Massive MIMO practically feasible in real deployments. Recently, the efficiency of combining massive MIMO and in-band wireless backhaul-based SC networks was studied in~\cite{li2015small, S2014if}, focusing on minimizing power consumption. The problem of user association for load balancing in heterogeneous networks (HetNets) has been studied in~\cite{2013user}. Although, users can be associated to more than one BS in order to reduce the load on the macro cell, deploying ultra-dense small cell networks makes user association more challenging. The work in~\cite{2013user} did not consider other important aspects in $5$G such as Massive MIMO, FD-enabled SCs, and mmWave communications. Recent work in~\cite{2015user} has addressed the user-cell association for Massive MIMO HetNets, which did not consider the joint optimization of load balancing, precoder design, and power allocation. Also the wireless backhaul faces the problem of limited-backhaul; hence the backhaul constraint needs to be considered.
Thus far, the key challenge of how to dynamically optimize the overall network performance taking into account the backhaul dynamics and constraints, and load balancing utilizing the combination of Massive MIMO, FD-enabled SCs, and mmWave communications has not been fully addressed~\cite{2015userSurvey}.

User association taking into account dynamic backhaul in $5$G HetNets faces a new challenge due to self-backhauled SCs, i.e., guaranteeing wireless backhaul capacity between MBS and SCs in order to offload the traffic from MBSs to SCs. It raises the following important question: Should MBS serve all macro UEs (MUEs) even though it is highly loaded or offload some MUEs to SCs subject to the wireless backhaul capacity? Due to the random deployment of massive number of devices, UEs around hotspots (i.e. airport lounges, shopping malls, stations, and other crowded places) may receive poor services from a-far-MBS with multiple beams focused on the same location. On the contrary, these UEs will receive better services from nearby SCs with a reliable wireless backhaul composed of strong single beam from the MBS or multiple received antennas at SCs.
\subsection{Main Contributions}
\label{Con}
The main contributions of this work are to study the problem of joint load balancing, interference mitigation, and in-band wireless backhauling taking into account dynamic backhaul and traffic load, which are listed as follows:
\begin{itemize}
  \item  The problem of joint load balancing (user association and user scheduling) and interference management (beamforming design and power allocation) for $5$G HetNets is modeled in which a DL scheduler is designed at the MBS to schedule macro UEs and provide backhaul to in-band FD-enabled SCs, with FD capability SCs serve both MUEs and small cell UEs in the same frequency band. Moreover, an interference management scheme is proposed to mitigate both co-tier and cross-tier interference from the MBS and FD-enabled SCs by designing a hierarchical precoding scheme and controlling the transmission of SCs. The problem is cast as a network utility maximization (NUM) problem subject to dynamic wireless backhaul constraints, traffic load, and imperfect channel state information (CSI). To make problem tractable, by invoking results from random matrix theory (RMT), we derive a closed-form expression of the signal-to-interference-plus-noise-ratio ($\mathrm{SINR}$) and transmit power when the numbers of MBS antennas and users grow very large.

  \item A Lyapunov framework is applied in order to solve the NUM problem in polynomial time. The NUM problem is decomposed into dynamic scheduling of MUEs, backhaul provisioning of FD-enabled SCs, and offloading MUEs to FD-enabled SCs. The joint load balancing and operation mode (FD or half-duplex) subproblem, which is a non-convex program with binary variables, is converted into a convex program by using the successive convex approximation (SCA) method. The motivations of using SCA are due to (i) its low complexity and fast convergence, and (ii) the obtained solution which yields many relaxed variables is close to zero or one.

  \item A performance evaluation is carried out to compare the proposed algorithm with other baselines under the impact of SCs density, number of BS antennas, and transmit power levels at low/high frequency bands. The effect of pilot training and channel aging is also studied to show the performance of Massive MIMO.

  \item A comprehensive performance analysis of our proposed algorithm based on the Lyapunov framework is provided. There exists an $[\mathcal{O}(1/\nu), \mathcal{O}(\nu)]$ utility-queue backlog tradeoff, which leads to an utility-delay balancing~\cite{vu2017ultra}, where $\nu$ is the Lyapunov control parameter. Moreover, a convergence analysis of the approximation method based on the SCA method is studied.
\end{itemize}
\subsection{Related Work}
\label{ReW}
The authors in~\cite{2015optimal} addressed the problem of dynamic resource control for HetNets with flexible backhaul (wired and wireless). However, the problem of load balancing when the number of antennas and users grows large is not considered. The user association problem has been studied for HetNets in~\cite{2013user, 2015user}, which does not take into account backhaul constraints. As pointed out in~\cite{2014overviewLB, 2015userSurvey} the current solutions for user association problem ignore the backhaul constraints, which is very crucial since the capacity of open access SCs with either wired or wireless backhaul always faces the limited backhaul constraint. Moreover, the load balancing problem should take into account imperfect CSI due to mobility, which is ignored in the previous work. Our previous work in~\cite{Vu2016} has considered the problem of joint in-band scheduling and interference mitigation in $5$G HetNets without considering the user association. In this work, we extend~\cite{Vu2016} by considering the load balancing problem taking into account the backhaul constraint and imperfect CSI, and further provides insights into the performance analysis of our proposed algorithm based on the Lyapunov framework and convergence of the SCA method.

The rest of this paper is organized as follows.\footnote{The lowercase letters, boldface lowercase letters, (boldface) uppercase letters and italic boldface uppercase letters are used to represent scalars, vectors, matrices, and sets, respectively. $\mathbf{X}^{\dag}$ and
$\text{rank}(\mathbf{X})$ denote the Hermitian transpose and the rank of matrix $\mathbf{X}$, respectively. $\text{diag}(x_1, x_2, ...x_\emph{N})$ denotes the block diagonal matric whose diagonal blocks are given by $x_1,
x_2, ..., x_\emph{N}$ and the identity matrix of size $N$ is denoted by $\mathbf{I}_\emph{N}$. The cardinality of a set $\mathcal{S}$,
is denoted  by $|\mathcal{S}|$. $\mathcal{CN}(0, \sigma^{2})$ denotes the Gaussian random distribution of zero mean and variance of $\sigma^{2}$.} Section~\ref{lb-SM-P} describes the system model and Section~\ref{Op-Form} provides the problem formulation for load balancing and interference mitigation. Section~\ref{LOF}
introduces the Lyapunov framework used to solve our problem. In Section~\ref{Evaluation}, we present the numerical results. We conclude the paper in Section~\ref{Conclusion}.
\section{System Model}
\label{lb-SM-P}
\subsection{System Model}
\label{lb-SystemModel}
\begin{figure}
    \centering
    \includegraphics[scale=0.5]{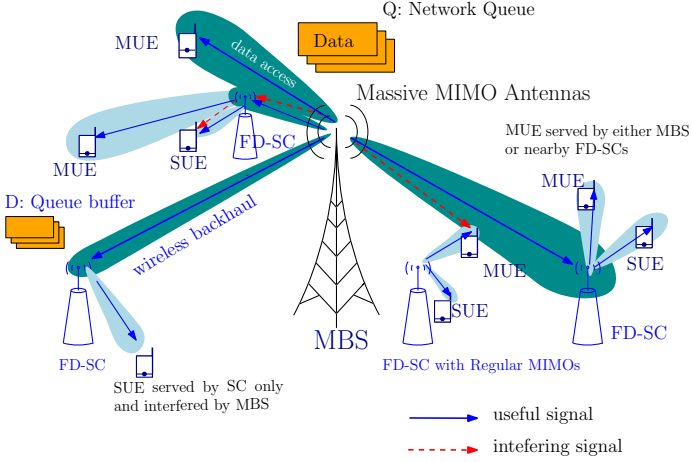}
    \caption{Integrated access and backhaul architecture for the considered 5G network scenario.}\label{DeploymentSetup}
\end{figure}

The downlink (DL) transmission of a HetNet scenario is considered as shown in Fig.~\ref{DeploymentSetup} in which a MBS $b_{0}$ is underlaid with a set of uniformly deployed $\emph{S}$ FD-enabled SCs,
$\mathcal{S} = \{b_{s}| s \in \{1, \ldots, \emph{S}\}\}$. Let $\mathcal{B} = \{b_{0}\} \cup \mathcal{S}$ denote the set of all base stations, where $|\mathcal{B}| = 1 +  \emph{S}$. The MBS is equipped with $\emph{N}$ number of antennas and serves a set of single-antenna $\emph{M}$ MUEs $\mathcal{M} = \{1, \ldots, \emph{M}\}$. Let $\mathcal{K} = \mathcal{M} \cup \mathcal{S}$ denote the set of users associated with MBS $b_0$, where $|\mathcal{K}| =  \emph{K} = \emph{M} +  \emph{S}$. The user indices $k = 1, 2, ..., M$ represent the corresponding MUEs indices $m = 1, 2, ..., M$, while the user indices $k = M+1, M+2, ..., M+S$ represent the corresponding SCs indices $s = 1, 2, ..., S$. We assume open access policy at FD-enabled SCs and each FD-enabled SC is equipped with $\Ns +1$ antennas: one receiving antenna is used for the wireless backhaul and $\Ns$ transmitting antenna to serve its single-antenna
small cell UEs (SUEs) or other MUEs at the same frequency band. Let $\mathcal{C}=\{ {\sue_{1}}, {\sue_{2}}, \ldots, {\sue_{S}} \}$ denote the set of SUEs, where $|\mathcal{C}| = \emph{S}$. Moreover, SCs are assumed to be FD capable with perfect
self-interference cancelation (SIC) capabilities\footnote{The case of imperfect SIC is left for future work.}. Co-channel time-division duplexing (TDD) protocol is considered in which the MBS and FD-enabled SCs share the entire bandwidth, and the DL transmission occurs at the same time. In this work, we consider a large number of antennas at both macro and SC BSs and a dense deployment of MUEs and SCs, such that $\emph{M},
\emph{N}, \Ns, \emph{S} \gg 1$.
\subsection{Channel Model}
\label{lb-ChannelModel}
We denote $\mathbf{h}_{m}^{(b_{0})} = \big[ h_{m}^{(b_{0},1)}, h_{m}^{(b_{0},2)}, \cdots, h_{m}^{(b_{0},\emph{N})} \big]^{T} \in \mathbb{C}^{\emph{N} \times 1}$ the propagation channel between the $m{\text{th}}$ MUE and the
antennas of the MBS $b_{0}$ in which $h_{m}^{(b_{0},n)}$ is the channel between the $m{\text{th}}$ MUE and the $n{\text{th}}$ MBS antenna. Let $\mathbf{H}^{(b_{0}), \text{M}} = \big[\mathbf{h}_{1}^{(b_{0})},
\mathbf{h}_{2}^{(b_{0})}, \cdots,
\mathbf{h}_{\emph{M}}^{(b_{0})} \big] \in \mathbb{C}^{\emph{N} \times \emph{M}}$ denote the channel matrix
between all MUEs and the MBS antennas. Moreover, we assume imperfect CSI for MUEs due to mobility and we denote
$\mathbf{\hat{H}}^{(b_{0}), \text{M}} = \big[ \mathbf{\hat{h}}_{1}^{(b_{0})}, \mathbf{\hat{h}}_{2}^{(b_{0})}, \cdots,
\mathbf{\hat{h}}_{\emph{M}}^{(b_{0})}\big] \in \mathbb{C}^{\emph{N} \times \emph{M}}$ as the estimate of
$\mathbf{H}^{(b_{0}), \text{M}}$ in which the imperfect CSI can be modeled as~\cite{rusek2013s}:
    \begin{equation} \label{Channel-Er}
    \mathbf{\hat{h}}_{m}^{(b_{0})} = \sqrt{\emph{N} \mathbf{\Theta}_{m}^{(b_{0})} } \mathbf{\hat{w}}_{m}^{(b_{0})},
    \end{equation}
where $\mathbf{\hat{w}}_{m}^{(b_{0})} = \sqrt{1 - {\tau_m}^2} \mathbf{w}_{m}^{(b_{0})} + \tau _m \mathbf{z}_{m}^{(b_{0})}$ is the estimate of the small-scale fading channel matrix and $\mathbf{\Theta}_{m}^{(b_{0})}$ is the
spatial channel correlation matrix that accounts for path loss and shadow fading. Here, $\mathbf{w}_{m}^{(b_{0})}$ and $\mathbf{z}_{m}^{(b_{0})}$
are the real channel and the channel noise, respectively, modeled as Gaussian random matrix with zero
mean and variance~$1/N$. The channel estimate error of MUE $m$ is denoted by $\tau _m, \tau_m \in [0, 1]$; in case of perfect CSI, $\tau _m = 0$. Similarly, let $\mathbf{H}^{(b_{0}), \text{S}} \in \mathbb{C}^{\emph{N} \times \emph{S}}$ and
$\mathbf{H}^{(b_{0}), \text{C}} \in \mathbb{C}^{\emph{N} \times \emph{S}}$ denote the channel matrices from the MBS antennas to SCs and SUEs,
respectively. Let $\mathbf{h}_{u}^{ {(b_s)}} \in \mathbb{C}^{\Ns \times 1}$ denote the channel propagation from SC $b_s$ to any receiver $u$. Let $c_s$ denote the SUE served by the SC $b_s$.
\section{Load Balancing and Interference Mitigation}
\label{Op-Form}
In this section, we formulate the joint optimization of user association, user scheduling, beamforming design, and power allocation. To that end, we first derive the received signal, data rate, and power transmit for each receiver (SCs are also treated as macro BS's UEs). We then formulate the problem as a network utility maximization subject to wireless backhaul constraints. However, the formulated problem does not have closed-form expressions for the objective and constraints. Hence, we apply RMT~\cite{wagner2012l} to get these closed-form expressions. We finally utilize the tool of stochastic optimization to decouple our problem into several solvable sub-problems.

The problem of user scheduling and user association for load balancing in the DL is addressed in which the MBS simultaneously provides data transmission to MUEs and wireless backhaul to the FD-enabled SCs, while the SCs with FD capability serve both SUEs and MUEs. For each MUE $m \in
\mathcal{M}$, let binary variable $\us_{m}^{(b_{s})}$ indicate the transmission association from BS $b_s \in \mathcal{B}$ to MUE $m$, i.e., $\us_{m}^{(b_{s})} = 1$ when MUE $m$ is associated with BS $b_s$, otherwise $\us_{m}^{(b_{s})} = 0$. Similarly, let binary variables $\us_{s+M}^{(b_{0})}$ and $\us_{c_s}^{(b_{s})}$ denote the transmission association indicators from MBS $b_0$ to SC $s$ and from SC $b_s$ to SUE $c_s$, respectively. We assume that each MUE $m$ connects to one BS (either MBS $b_0$ or SC $b_s$) at time slot $t$. Each SC is equipped with $\Ns$ transmitting
antennas, and we assume that each SC serves up to $\emph{N}_{s}^{\text{au}}$ active users (either SUE or MUE) at each time slot, such that $\emph{N}_{s}^{\mathrm{au}} \leq {\Ns}$, where the superscript $\mathrm{au}$ stands for ``active users". Hence, we have the following constraints for load balancing:
\begin{equation} \label{eq:association1}
 \textstyle  \sum _{s = 0}^{\emph{S}} \us_{m}^{(b_{s})} \leq 1, \textstyle \sum _{m = 1}^{\emph{M}} \us_{m}^{(b_{s})}  + \us_{c_s}^{(b_{s})} \leq \emph{N}_{s}^{\mathrm{au}}, \forall~s, m \in \mathcal{K}.
\end{equation}
We define vector $\mathbf{\us} = \big \{\us_{j}^{(b_s)}| b_s \in \mathcal{B}, j \in \{\mathcal{M} \cup \mathcal{S} \cup \mathcal{C}\} \big \}$ containing all transmission indicators between BSs and UEs. Let
$\emph{N}_{s}^{\mathrm{tx}} = \sum _{m = 1}^{\emph{M}} \us_{m}^{(b_{s})}  + \us_{c_s}^{(b_{s})}$ be the total number of transmissions at SC, where superscript $\mathrm{tx}$ stands for ``transmissions", and thus the latter of
(\ref{eq:association1}) becomes $\emph{N}_{s}^{\mathrm{tx}} \leq \emph{N}_{s}^{\mathrm{au}}, \forall s \in \mathcal{S}$.
\subsection{Downlink Transmission Signal} 
\label{DLTX}
The MBS serves two types of users: MUEs with imperfect CSI and FD-enabled SCs with perfect CSI. Let $p_{m}^{(b_0)}$, $p_{s+M}^{(b_0)}$, and $P^{(b_0)}$ denote the DL MBS transmit power assigned to MUE $m$, the DL MBS transmit power assigned to SC $s$, and the maximum transmit power at the MBS, respectively. We focus on the multiple-input single-output (MISO) channel, where the MBS with $\emph{N}$ antennas can serve $\emph{K}$ UEs. Here, we take into account user scheduling and association, and our proposal can apply to any special case when number of UEs is larger than number of antennas, i.e., $\emph{K} > \emph{N}$. SC exploits FD capability to double capacity, FD-enabled SC causes unwanted FD interference: cross-tier interference to adjacent MUEs (or other SCs), and co-tier interference to other UEs. Hence, in order to convert the interference channel to the MISO channel, we design a precoder at the MBS and propose an operation mode policy to control FD interference in order to treat the total FD interference as additional noise.

\begin{definition}\label{USP}[Operation Mode Policy] We define $\boldsymbol{\op}$ as the operation mode to control the FD-enabled SC transmission to reduce FD interference. The operation mode is expressed as $\boldsymbol{\op}(t) = \{ \op^{(b_s)}(t)~|~ \op^{(b_s)}(t) \in \{0, 1\}, \forall s \in \mathcal{S}  \}$. Here, $\op^{(b_s)}(t) = 1$ indicates SC $b_{s}$ operates in FD mode and $\op^{(b_s)}(t) = 0$ for half-duplex (HD) mode.
\end{definition}

We assume that the MBS uses a precoding scheme, $\mathbf{V} = [\mathbf{v}_1, \mathbf{v}_2, \ldots, \mathbf{v}_{\emph{K}}] \in \mathbb{C}^{\emph{N} \times \emph{K}}$. To exploit the  degrees of freedom of massive MIMO, the hierarchical interference mitigation scheme in~\cite{Liu2014, 2015RZF} is applied to design the
precoder, i.e., $\mathbf{V} = \mathbf{U} \mathbf{T}$, where $\mathbf{T} \in \mathbb{C}^{\emph{N} \times \emph{N}_{\mathrm{itf}}}$  is used to control co-tier interference and capture the spatial multiplexing gain, and $\mathbf{U} \in \mathbb{C}^{\emph{N}_{\mathrm{itf}} \times \emph{K}}$ is used to suppress cross-tier interference. Here, $\emph{N}_{\mathrm{itf}} < \emph{N}$, where the subscript $\mathrm{itf}$ stands for
``interference". The precoder $\mathbf{U}$ is chosen such that
\begin{equation}\label{zero-ICI}
\mathbf{U}^{\dag} \textstyle \sum _{s = 1}^{\emph{S}} \op^{(b_s)} \mathbf{\Theta}_{{s}}^{(b_0)} = 0,
\end{equation}
where $\mathbf{\Theta}_{{s}}^{(b_0)} \in \mathbb{C}^{\emph{N} \times \emph{N} }$ is the sum of the correlation matrices between MBS antennas and users belong to SC ${s}$. Here, $\mathbf{U}$ is in the null space of $\sum  ^S_{s=1} \beta^{(b_s)} \mathbf{\Theta}_{{s}}^{(b_0)}$. Note that $\op^{(b_s)}$ determines
that the transmission of FD-enabled SC is enabled or not. The precoder $\mathbf{T}$ is designed to adapt to the real time CSI based on $\mathbf{\hat{H}^{\dag} \mathbf{U} } \in \mathbb{C}^{\emph{K} \times \emph{N}_{\mathrm{itf}}}$,
where $\mathbf{\hat{H}} = [ \mathbf{\hat{h}}^{(b_{0})}]_{k \in\mathcal{K}}^{\dag}$. In this paper, we consider the regularized zero-forcing (RZF) precoding\footnote{Other precoders are left for future work.} that is given by
$\mathbf{T} = \big(\mathbf{U}^{\dag} \mathbf{\hat{H}}^{\dag}  \mathbf{\hat{H}}\mathbf{U} + \emph{N} \alpha \mathbf{I}_{\emph{N}_{\mathrm{itf}}}\big)^{-1}  \mathbf{U}^{\dag} \mathbf{\hat{H}}^{\dag}$, where the regularization
parameter $\alpha > 0$ is scaled by $\emph{N}$ to ensure that the matrix $\mathbf{U}^{\dag} \mathbf{\hat{H}}^{\dag} \mathbf{\hat{H}}\mathbf{U} + \emph{N} \alpha \mathbf{I}_{\emph{N}_{\mathrm{itf}}}$ is well conditioned as
$\emph{N} \rightarrow \infty$. The precoder $\mathbf{T}$ is chosen to satisfy the power constraint $\text{Tr}\big (\mathbf{P} \mathbf{T}^{\dag} \mathbf{T} \big) \leq P^{(b_0)}$, where $\mathbf{P} = \text{diag}(p_1^{(b_0)},
p_2^{(b_0)}, \ldots, p_{\emph{K}}^{(b_0)})$. We also assume that each SC uses  ZF precoding  to server its users, ${\mathbf{F}}^{(b_s)} = [\mathbf{f}_{1}^{(b_s)}, \mathbf{f}_2^{(b_s)}, \ldots, \mathbf{f}_{\emph{N}_{s}^{\text{tx}}}^{(b_s)}] \in \mathbb{C}^{\Ns \times \emph{N}_{s}^{\text{tx}}}$ which reads $\mathbf{f}_u^{(b_s)} = \mathbf{h}_u^{(b_s)\dag} \big( \mathbf{h}_u^{(b_s)}\mathbf{h}_u^{(b_s)\dag}\big)^{-1}$ such that $\mathbf{F}^{(b_s)}$ is chosen to satisfy the equality power constraint $\text{Tr}\big (\mathbf{P}^{(b_s)} \mathbf{F}^{(b_s)\dag} \mathbf{F}^{(b_s)} \big) = P^{(b_s)}$\footnote{We choose the equality constraints for transmit power at SCs to reach the optimal rate at maximum power rather than using $\text{Tr}\big (\mathbf{P}^{(b_s)} \mathbf{F}^{(b_s)\dag} \mathbf{F}^{(b_s)} \big) \leq P^{(b_s)}$, since the power at SCs is relatively small.}. Here, $\mathbf{P}^{(b_s)} = \text{diag}(p_1^{(b_s)},
p_2^{(b_s)}, \ldots, p_{ \emph{N}_{s}^{\text{tx}}} ^{(b_s)})$. The channel propagation from the SC $b_s$ to the MUE $m$ (referred to as user $u$) is $\mathbf{h}_u^{(b_s)} = \mathbf{\hat{h}}_m^{(b_s)} = \sqrt{ N_s \mathbf{\Theta}_{m}^{(b_{s})} } \big( \sqrt{1 - {\tau_m}^2} \mathbf{w}_{m}^{(b_{s})} + \tau _m \mathbf{z}_{m}^{(b_{s})} \big)$, where $\mathbf{\Theta}_{m}^{(b_{s})}\in \mathbb{C}^{N_s \times N_s}$ is the channel correlation matrix. Here, $\mathbf{w}_{m}^{(b_{s})}$ and $\mathbf{z}_{m}^{(b_{s})}$ are the real channel and the channel noise from SC $b_s$ to MUE $m$, respectively, modeled as a Gaussian random matrix with zero mean and variance~$1/N_s$.

By utilizing a massive number of antennas at MBS, a large spatial degree of freedom is utilized to serve MUEs
and FD-enabled SCs, while the remaining degrees of freedom are used to mitigate cross-tier interference. In massive MIMO system, the total number of antennas is considered as the degree of freedom~\cite{Liu2014}. Hence, we have the antenna constraint for user association and operation mode such that $\sum_{k =
1}^{\emph{K}} \us_{k}^{(b_0)}(t) + \sum_{s =  1}^{\emph{S}} \emph{N}_{s}^{\text{tx}}(t) \leq \emph{N}$. For notational simplicity, we remove the time dependency from the symbols throughout the discussion. The received signal
$y_{m}^{(b_0)}$  at each MUE $m \in \mathcal{M}$ at time instant $t$ is given by
\begin{equation}\label{SINR-MUE-1} %
\begin{split}
\displaystyle y_{m}^{(b_0)} =~& \us_{m}^{(b_0)} \sqrt{p_{m}^{(b_0)}} \mathbf{h}_{m}^{(b_0) \dag} \mathbf{v}_{m} x_{m}^{(b_0)}
\\& + \underbrace{\textstyle \sum \limits_{s = 1}^{\emph{S}} \op^{(b_s)}  \textstyle \sum _{u =
1}^{\emph{N}_{s}^{\mathrm{tx}}} \us_{u}^{(b_s)} \sqrt{\pc_{u}^{(b_s)}} \mathbf{h}_{m}^{(b_s) \dag} \mathbf{f}_{u}^{(b_{s})}  {x}_{u}^{(b_s)}}_{\text{cross-tier
interference}} \\& +  \underbrace{ \textstyle \sum \limits_{k = 1, k \neq m}^{\emph{K} } \us_{k}^{(b_0)}\sqrt{p_{k}^{(b_0)}} \mathbf{{h}}_{m}^{(b_0) \dag} \mathbf{v}_{k} x_{k}^{(b_0)}}_{\text{co-tier interference}} +
\eta_{m},
\end{split}
\end{equation}
where $x_{m}^{(b_0)}$ is the signal symbol from the MBS to the MUE $m$, $\mathbf{v}_{m}$ is the precoding vectors of MUE $m$, and $\eta_{m} \sim \mathcal{CN}(0,1)$ is the
thermal noise at MUE $m$. While ${x}_{u}^{{(b_s)}}$ is the transmit signal symbol from SC $b_{s}$ to its user $u$.

At time instant $t$, the received signal $y_{s+M}^{(b_0)}$ at each SC $s \in \mathcal{K}$ suffers from self-interference, cross-tier and co-tier interference, which is given by
\begin{equation} \label{SINR-SC-1}
\begin{split}
y_{s+M}^{(b_0)}  =~& \us_{s+M}^{(b_0)} \sqrt{p_{s+M}^{(b_0)}}  \mathbf{h}_{s+M}^{(b_0) \dag} \mathbf{v}_{s+M}^{\text{}} x_{s+M}^{(b_0)} \\& + \underbrace{\textstyle \sum \limits_{s' = 1, s' \neq s}^{\emph{S}} \op^{(b_{s'})} \textstyle \sum
\limits_{u' = 1}^{\emph{N}_{s'}^{\text{tx}}} \us_{u'}^{(b_{s'})} \sqrt{\pc_{u'}^{(b_{s'})}} \mathbf{h}_{s}^{ (b_{s'}) \dag}  \mathbf{f}_{u'}^{(b_{s'})} {x}_{{u'}}^{ (b_{s'}) }}_{\text{cross-tier interference}}
\\& + \underbrace{ \op^{(b_s)} \textstyle \sum \limits_{u = 1}^{\emph{N}_{s}^{\mathrm{tx}}} \us_{u}^{(b_s)} \sqrt{\pc_{u}^{(b_s)}} \mathbf{h}_{s}^{ {(b_s)} \dag} \mathbf{f}_{u}^{(b_{s})} {x}_{u}^{ (b_s)
}}_{\text{self-interference}} \\& + \underbrace{\textstyle \sum \limits_{k = 1, k \neq s+M}^{\emph{K} } \us_{k}^{(b_0)} \sqrt{p_{k}^{(b_0)}} \mathbf{h}_{s+M}^{ (b_0) \dag} \mathbf{v}_{k} x_{k}^{ (b_0)} }_{\text{co-tier interference}} + \eta_{s+M},
\end{split}
\end{equation}
where $x_{s+M}^{(b_0)}$ is the signal symbol from the MBS to the SC $s$, $\mathbf{v}_{s+M}$ is the precoding vectors of SC $s$, and $\eta_{s+M} \sim \mathcal{CN}(0,1)$ is the thermal noise of the SC ${s}$.

The received signal from the SC $b_{s}$ at receiver $u$, $y_{u}^{(b_s)} = 0$, if the SC $b_{s}$ operates in HD mode, $\op^{(b_s)} = 0$. For FD mode, $\op^{(b_s)} = 1$, the received signal $y_{u}^{(b_s)}$ is given by
\begin{equation} \label{SINR-Sue-1}
\begin{split}
y_{u}^{(b_s)} =~&  \op^{(b_s)}  \us^{(b_{s})}_{u} \sqrt{\pc_{u}^{(b_s)}} \mathbf{h}_{u}^{(b_s) \dag } \mathbf{f}_{u}^{(b_s)} x_{u}^{(b_s)}  \\&  + \underbrace{\textstyle \sum \limits_{s' = 1, s' \neq s}^{\emph{S} } \op^{(b_{s'})}  \textstyle
\sum \limits_{u' = 1, }^{\emph{N}_{s'}^{\mathrm{tx}}} \us_{u'}^{(b_{s'})} \sqrt{\pc_{{u'}}^{(b_{s'})}}  \mathbf{h}_{u}^{(b_{s'}) \dag} \mathbf{f}_{u'}^{(b_s')} x_{{u'}}^{(b_{s'})} }_{\text{co-tier interference}}  \\
& + \underbrace{ \op^{(b_s)}  \textstyle \sum \limits_{j = 1, j \neq u}^{\emph{N}_{s}^{\text{tx}}} \us_{j}^{(b_{s})} \sqrt{\pc_{j}^{(b_s)}} \mathbf{h}_{u}^{(b_s) \dag } \mathbf{f}_{j}^{(b_s)} x_{u}^{(b_s)} }_{\text{co-tier
self-interference}} \\&  + \underbrace{ \textstyle \sum \limits_{k = 1, k \neq u}^{\emph{K} } \us_{k}^{(b_0)} \sqrt{p_{k}^{(b_{0})}} \mathbf{h}_{u}^{ (b_{0}) \dag} \mathbf{v}_{k} x_{k}^{(b_{0})} }_{\text{cross-tier interference}}
+ \eta_{u},
\end{split}
\end{equation}
where $x_{u}^{(b_s)}$ is the transmit data symbol from the SC $b_{s}$ to receiver $u$ and $\eta_{u} \sim \mathcal{CN}(0,1)$ is the thermal noise at receiver $u$. We imply that the receiver $u$ can be either a SUE or an MUE.

The precoder $\mathbf{V}$ is designed at the MBS to null the co-tier interference and to remove completely the cross-tier interference to SCs's users~(\ref{zero-ICI}) and the self-interference is well treated, while $\text{Tr}\big (\mathbf{P}^{(b_s)} \mathbf{F}^{(b_s)\dag} \mathbf{F}^{(b_s)} \big) = P^{(b_s)}$. Thus, according to~(\ref{SINR-MUE-1})-(\ref{SINR-Sue-1}), the $\mathrm{SINRs}$ of an MUE $m$ served by MBS, a SC ${s}$ served by MBS, a receiver $u$ served by SC are given in ~\eqref{SINR-MUE-2}-\eqref{SINR-SUE-2}, respectively.

\begin{floatEq}
\begin{equation} \label{SINR-MUE-2}
\begin{split} \gamma_m^{(b_0)} =  \frac { \us_{m}^{(b_0)} p_{m}^{(b_0)} |\mathbf{h}_{m}^{(b_0) \dag} \mathbf{v}_{m} |^2 }  {  {\textstyle \sum _{k \neq m} \us_{k}^{(b_0)} p_{k}^{(b_0)} | \mathbf{h}_{m}^{(b_0) \dag}
\mathbf{v}_{k}|^2  + \textstyle \sum_{s} \op^{(b_{s})} P^{(b_s)} |\mathbf{h}_{m}^{(b_s) \dag}|^2 + 1}}.
\end{split}
\end{equation}
\begin{equation} \label{SINR-SC-2}
\begin{split}
\gamma_{s+M}^{(b_0)} =  \frac { \us_{s+M}^{(b_0)} p_{s+M}^{(b_0)} | \mathbf{h}_{s+M}^{(b_0) \dag} \mathbf{v}_{s+M}|^2 }{  {\textstyle \sum _{k \neq
s+M} \us_{k}^{(b_0)} p_{k}^{(b_0)} |\mathbf{h}_{s+M}^{(b_0) \dag} \mathbf{v}_{k}|^2  + \textstyle \sum _{s' \neq s} \op^{(b_{s'})} P^{(b_{s'})} |\mathbf{h}_{s}^{(b_{s'}) \dag}|^2  + 1 }}.
\end{split}
\end{equation}
\begin{equation} \label{SINR-SUE-2}
\begin{split}
\gamma_{{u}}^{(b_s)} =   \frac {\op^{(b_s)}  \us^{(b_{s})}_{u} \pc_u^{(b_s)} |\mathbf{h}_{{u}}^{(b_s) \dag}  \mathbf{f}_{u}^{(b_s)}|^2} {  \op^{(b_s)}  \textstyle \sum _{j = 1, j \neq u}^{ } \us_{j}^{(b_{s})} {\pc_{j}^{(b_s)}} |
\mathbf{h}_{u}^{(b_s) \dag } \mathbf{f}_{j}^{(b_s)} |^2 + {\textstyle \sum _{s' \neq s} \op^{(b_{{s'}})} P^{(b_{s'})} |\mathbf{h}_{{u}}^{ {(b_{s'}) } \dag} |^2  + 1}}.
\end{split}
\end{equation}
\end{floatEq}

\subsection{Joint Load Balancing and Interference Mitigation Algorithm}
\label{NUM-Pro}
Let us consider a joint optimization of load balancing $\mathbf{l}$, operation mode $\boldsymbol{\op}$, interference mitigation $\mathbf{U}$, and transmit power allocation $\mathbf{p} = (p_{1}^{(b_0)}, p_{2}^{(b_0)}, \ldots,
p_{\emph{K}}^{(b_0)})$ that satisfies the transmit power budget of MBS i.e. , $\text{Tr}\big (\mathbf{P} \mathbf{T}^{\dag} \mathbf{T} \big) \leq P^{(b_0)}$. We define $\EP_{k}^{(b_s)} = \frac{ P^{(b_s)}
|\mathbf{h}_{k}^{(b_s)
\dag}|^2 }{ |\eta_k|^2}$ and $\epsilon_{o}$ as the FD interference to noise ratio (INR) from FD-enabled SC $b_{s}$ to any scheduled receiver
$k$, and the allowed FD INR threshold, respectively. The FD interference threshold is defined such that $\textstyle \sum_{k=1}^{\emph{K}} \textstyle \sum_{s = 1}^{\emph{S}}\EP_{k}^{(b_s)} \leq \epsilon_{o}$, such that the total FD
interference is considered as noise. Under the operation mode policy, we schedule the receiver $i$ and enable the transmission of SC $b_s$ as long as $\textstyle \sum_{k=1}^{\emph{K}} \textstyle \sum_{s =1}^{\emph{S}}
\us_{k}^{(b_0)} \op^{(b_s)} \EP_{k}^{(b_s)} \leq \epsilon_{o}$. Let $\mathbf{\Lambda}^{o} = \{ \mathbf{\us}, \boldsymbol{\op}\}$ be a composite control variable of user association and operation mode. We define
$\mathbf{\Lambda} = \{ \mathbf{\Lambda}^{o}, \mathbf{U}, \mathbf{p}\}$ as a composite control variable, which adapts to the spatial channel correlation matrix $\mathbf{\Theta}$.

For a given $\mathbf{\Lambda}$ that satisfies~(\ref{zero-ICI}) and operation mode policy, the respective Ergodic data rates of SC $s$ and SUE $u$ are $ r_{s+M} ( \mathbf{\Lambda} | \mathbf{\Theta} )  = \mathbb{E} \big[ \log \big( 1 +  \gamma_{s+M}^{ (b_0) }  \big) \big]$ and $r_{u}^{(b_s)}(\mathbf{\Lambda}|\mathbf{\Theta})  =  \mathbb{E} \big[ \log
 \big( 1 + \gamma_{u}^{(b_s)}  \big) \big]$. While from the constraint~(\ref{eq:association1}) the Ergodic data rate of MUE $m$ will depend on which BS the MUE is associated with, i.e., $r_{m}(\mathbf{\Lambda}|\mathbf{\Theta})  =  \mathbb{E} \big[  \log \big( 1 + \gamma_m^{(b_0)} \big) \big] +   \sum\limits_{s=1}^{\emph{S}}\min \{ \mathbb{E} \big[ \log \big( 1 + \gamma_m^{(b_s)}\big) \big], \quad r_{s} ( \mathbf{\Lambda} | \mathbf{\Theta} ) - \sum\limits_{u \neq m} r_{u}^{(b_s)}(\mathbf{\Lambda}|\mathbf{\Theta}) \} $. In other words, the first term is the data rate from from the MBS to MUE when MUE is associated with the MBS, while the second term is when the FD-enabled SCs allow MUE to connect (If MUE is connected to the FD-enabled SC, then the rate of MUE should be the minimum between $r_{m}^{(b_s)}(\mathbf{\Lambda}|\mathbf{\Theta})$ and data stream from the MBS via FD-enabled SC to MUE, excepts other SC's users).
\begin{definition}  For any vector $\mathbf{x}(t) = ({x}_{1}(t), ...,{x}_{\emph{K}}(t))$, let $\bar{\mathbf{x}} =
(\bar{x}_{1}, \cdots,\bar{x}_{\emph{K}})$ denote the time average expectation of $\mathbf{x}(t)$, where $\textstyle \bar{\mathbf{x}}
\triangleq \lim_{t \to \infty} \frac{1}{t}  \sum_{\tau=0}^{t-1} \mathbb{E}[\mathbf{x}(\tau)]$. Similarly, $\textstyle \bar{\mathbf{r}} \triangleq \lim_{t \to \infty} \frac{1}{t}  \sum_{\tau=0}^{t-1}
\mathbb{E}[\mathbf{r}(\tau)]$ denotes the time average expectation of the Ergodic data rate.
\end{definition}
For a given composite control variable $\mathbf{\Lambda}$ that adapts to the spatial channel correlation matrix $\mathbf{\Theta}$, the average data rate region is defined as the convex hull of the average data rate of users,
which is expressed as:
\begin{equation}\label{eq:rate-region}
\begin{split}
\textstyle \mathcal{R} \triangleq ~&\bigg \{ \bar{\mathbf{r}}(\mathbf{\Lambda}|\mathbf{\Theta}) \in \mathbf{R}^{\emph{K}}_{+}~|~\mathbf{\us} \in \{0, 1\}^{\emph{K} + \emph{MS} + \emph{S}},  \boldsymbol{\op} \in \{0, 1\}^{\emph{S}},\\& \textstyle \sum_{s = 0}^{\emph{S}} \us_{m}^{(b_{s})} \leq 1,\quad\forall ~m \in \mathcal{M},
\\&  \textstyle \sum_{m = 1}^{\emph{M}} \us_{m}^{(b_{s})}  + \us_{c_s}^{(b_{s})} = \emph{N}_{s}^{\mathrm{tx}}, \emph{N}_{s}^{\text{tx}} \leq \emph{N}_{s}^{\mathrm{au}},\quad\forall ~b_s \in \mathcal{S},
\\& \textstyle \sum_{k = 1}^{\emph{K}} \us_{k}^{(b_0)} + \textstyle \sum_{s =  1}^{\emph{S}} \emph{N}_{s}^{\mathrm{tx}}  \leq \emph{N},
\\& \textstyle \sum_{k= 1 }^{\emph{K}} \textstyle \sum_{s = 1}^{\emph{S}} \us_{k}^{(b_0)} \op^{(b_s)} \EP_{k}^{(b_s)} \leq \epsilon_{o},
\\& \text{Tr}\big (\mathbf{P} \mathbf{T}^{\dag} \mathbf{T} \big) \leq P^{(b_0)},~\mathbf{U}^{\dag} \textstyle \sum _{s = 1}^{\emph{S}} \op^{(b_{s})} \mathbf{\Theta}_{s}^{(b_0)} = 0  \bigg \},\nonumber
\end{split}
\end{equation}
where $\bar{\mathbf{r}}(\mathbf{\Lambda}|\mathbf{\Theta}) = (\bar{r}_{1}(\mathbf{\Lambda}|\mathbf{\Theta}), \ldots, \bar{r}_{K}(\mathbf{\Lambda}|\mathbf{\Theta}))^{T}$. Following the results from~\cite{pareto2011}, the
boundary points of the rate regime with total power constraint and no self-interference are Pareto-optimal\footnote{The Pareto optimal is the set of user rates at which it is impossible to improve any of the rates without
simultaneously decreasing at least one of the others.}. Moreover, according to~\cite[Proposition 1]{pareto2012}, if the INR covariance matrices approach the identity matrix, the Pareto rate regime of the MIMO interference
system is convex. Hence, our rate regime is a Pareto-optimal, and thus is convex with above constraints.

Let us assume that each FD-enabled SC acts as a relay to forward data to its users. If the MBS transmits data to FD-enabled SC $b_s$, but the transmission of SC $b_s$ is disabled, it cannot serve its SUE. Hence, we define $\mathbf{D(t)} = (D_1(t), D_2(t), \ldots, D_{\emph{S}}(t))$ as a data queue at SCs, where at each time slot $t$,
the wireless backhaul queue at FD-enabled SC $b_s$ is
\begin{equation}\label{queueD0}
{D}_{s}(t + 1) = \mathrm{max}[{D}_{s}(t) + {r}_{s+M}(t) -   {r}_{\sue_{s}}^{(b_s)}(t), \quad 0], \quad\forall~s \in \mathcal{S}.
\end{equation}
The SC offloads some MUEs from the MBS if the wireless backhaul capacity between the SCs and the MBS is guaranteed, and hence, for each SC we have the following wireless backhaul condition for all $t \geq 0$: ``If the access link between the MUE $m$ and the MBS is better than the link between the MUE $m$ and the SCs, then the MUE connects with the MBS rather than with other SCs", i.e.,\footnote{The queues of MUEs are handled at the MBS and SCs strictly handle data for SUEs, hence when SCs open connection for MUEs, they should have immediate capacity in terms of data rate. We do not include the constraint~(\ref{UA-Condition}) for the closed access case in~\cite{Vu2016}.}
\begin{equation}\label{UA-Condition}
\text{if} \quad  {r}_{s+M}^{}(t)  \leq {r}_{m}^{(b_0)}(t), \quad \text{then} \quad \l_m^{(b_s)} = 0, \quad \forall s \in \mathcal{S}, m \in \emph{K}.
\end{equation}

\begin{definition}\label{QueueDef}[Queue stability] For any discrete queue ${Q}(t)$ over time slots $t \in \{0, 1, \ldots\}$ and ${Q}(t) \in R_{+}$, ${Q}(t)$ is stable if $\textstyle \bar{Q}
\triangleq \lim_{t \to \infty} \frac{1}{t}  \sum_{\tau=0}^{t-1} \mathbb{E} \big [|{Q}(\tau)| \big ] < \infty$. A queue network is stable if each queue is stable.
\end{definition}
We define the network utility function ${f}_0 (\cdot)$ to be non-decreasing, concave over the convex region $\mathcal{R}$ for a given $\mathbf{\Theta}$. The objective is to maximize the network utility under wireless
backhaul constraints and imperfect CSI. Thus, the NUM problem is given by,
\begin{subequations}
\label{eq:Obj-Formulate-0}
\begin{align}
    \underset{\bar{\mathbf{r}}}{\text{max}}
        & \quad {f}_{0} (\bar{\mathbf{r}}) \label{eq-01}\\
    \text{subject to}
        & \quad  \eqref{UA-Condition}, \quad \bar{\mathbf{r}} \in \mathcal{R}, \quad \bar{\mathbf{D}} < \infty,\label{eq-02}
\end{align}
\end{subequations}
where ${f}_0 ( \bar{\mathbf{r}} ) =   \textstyle \sum _{k = 1}^{\emph{K}} \omega_{k}(t) f( \bar{r}_{k} )$ with $\omega_{k}(t) \geq 0$ is the weight of user $k$, $f(\cdot)$ is assumed to be twice differentiable, concave,
and increasing $L$-Lipschitz function for all $\bar{r} \geq 0$. Solving~(\ref{eq:Obj-Formulate-0}) is non-trivial since the average rate region $\mathcal{R}$ does not have a tractable form. To overcome this challenge, we need to find closed-form expressions of the data rate and the average transmit power. Inspired by~\cite{wagner2012l}, we invoke RMT to get the closed-form expressions for the user data rate and transmit power as $\emph{N} \gg \emph{K}$.
\subsection{Closed-Form Expression via Deterministic Equivalent}
\label{DL-Signal}
We invoke recent results from RMT in order to get the deterministic equivalent of user rate and transmit power via $\textbf{Theorem~\ref{Theorem1}}$.
\begin{theorem}\label{Theorem1} Recall that $\alpha$ is the RZF parameter. As ${N} \gg  {K}$; ${N}, {K} \rightarrow \infty$, by applying the technique in~\cite[Theorem 2]{wagner2012l}, the deterministic equivalent of the asymptotic $\mathrm{SINR}$ of  MUE $m$ is
	\begin{equation}
	\gamma_{m}^{(b_0)} \xrightarrow{a. s.} \frac { \us_{m}^{(b_0)} p_{m}^{(b_0)} (1 - \tau_m^2) (\Omega_{m}^{})^2}{ \Phi }, \nonumber
	\end{equation}
	where $\xrightarrow{a.s.}$ denotes the almost sure convergence and $\Phi = \Upsilon_{m}^{}\Big[ \alpha^2 -  \tau_m^2 \big(\alpha^2 - (\alpha + \Omega_{m})^2 \big)\Big] +(\alpha + \Omega_{m}^{})^2(1 + \textstyle \sum _{s = 1}^{\emph{S}} \op^{(b_s)} \EP_{m}^{(b_s)})$. Here, $\Omega_{m} =  \frac{1}{\emph{N}} \text{Tr} (\mathbf{\tilde{\Theta}}_m \mathbf{G})$ forms the unique positive solution of
	which is the Stieltjes transform of nonnegative finite measure~\cite[Theorem 1]{wagner2012l}, where  $\textstyle \mathbf{G} =  \Big( \frac{1}{\emph{N}} \sum _{k = 1}^{\emph{K}} \frac{ \mathbf{\tilde{\Theta}}_k }{ \alpha +
		\Omega_{k}^{}} + \mathbf{I}_{\emph{N}_{\text{itf}}} \Big)^{-1}$. In addition, $\textstyle \Upsilon_{m} =  \frac{1}{\emph{N}} \sum _{k = 1, k \neq m}^{\emph{K}} \frac{ \alpha^2 \us_{k}^{(b_0)} p_{k}^{(b_0)} e_{km}}{
		(\alpha + \Omega_{k}^{})^2 }$, and  $\mathbf{\tilde{\Theta}}_k = \mathbf{U} \mathbf{U}^{\dag} \mathbf{\Theta}_k^{(b_0)} \mathbf{U} \mathbf{U}^{\dag}$. $\mathbf{e} = [e_{k}], k \in \mathcal{K}$, and $\mathbf{e_m} = [e_{mk}], k \in \mathcal{K}$ are given
	by
	$\mathbf{e} = ( \mathbf{I} - \mathbf{J})^{-1}\mathbf{u}$, $\mathbf{e_k} = ( \mathbf{I} - \mathbf{J})^{-1}\mathbf{u_k}$, where $\mathbf{J} = [J_{ij}], i,j \in \mathcal{K}$. $\mathbf{u} = [u_{k}], k \in \mathcal{K}$,
	$\mathbf{u_m} = [u_{mk}], k \in \mathcal{K}$ are given by $\mathbf{J}_{ij} = \displaystyle  \frac{ \frac{1}{\emph{N}} \text{tr} { \mathbf{\tilde{\Theta}}_i \mathbf{G} \mathbf{\tilde{\Theta}}_j \mathbf{G}} } {\emph{N} (\alpha
		+ \Omega_{j}^{})^2}$, $u_{mk} = \frac{1}{\alpha^2 N} \text{tr} { \mathbf{\tilde{\Theta}}_k \mathbf{G}  \mathbf{\tilde{\Theta}}_m \mathbf{G}}$, $u_{k} = \frac{1}{\alpha^2 \emph{N}} \text{tr} { \mathbf{\tilde{\Theta}}_k
		\mathbf{G}^2}$. Similarly, the SINR of  SC $b_s$ is
	\begin{equation}
	\gamma_{s}^{(b_0)} \xrightarrow{a.s.}  \frac { \us_{s}^{(b_0)} p_{s}^{(b_0)} (\Omega_{s}^{})^2}  { \alpha^2
		\Upsilon_{s}^{} + (\alpha + \Omega_{s}^{})^2(1 + \sum _{s' = 1, s' \neq s}^{\emph{S}} \op^{(b_{s'})} \EP_{s}^{(b_{s'})}) }. \nonumber
	\end{equation}
The power constraint at the MBS can be calculated as  $\frac{1}{N} \sum _{k = 1}^{\emph{K}} \frac{ {p}_{k}^{(b_0)} \alpha^2 e_{k}}{ (\alpha + \Omega_{k}^{})^2 } - P^{(b_0)} \leq 0$. Moreover, following the analysis in the proof of~\cite[Theorem 3]{wagner2012l}, \cite[Lemma 6]{Liu2014} for a small fixed\footnote{The deterministic equivalent holds for a small fixed $\alpha$ as studied in \cite{Liu2014}, while the problem of finding the optimal value $\alpha$ has been studied in \cite{wagner2012l, 2015RZF}.} $\alpha > 0$, $\Upsilon_{k}^{} = \mathbi{O}(1)$ and $\alpha^2 e_{k} = \Omega_{k}^{} + \mathbi{O}(\alpha)$ yield  the deterministic equivalent of the asymptotic SINRs of UEs (\ref{SINR-MUE-2})-(\ref{SINR-SUE-2}) as
\begin{equation}\label{SINR-MUE-3}
\textstyle \gamma_{m}^{(b_0)}(\mathbf{\Lambda}|\mathbf{\Theta}) \xrightarrow{a.s.} \textstyle \frac{\us_m^{(b_0)} p_{m}^{(b_0) } (1 - \tau_m^2)}{1 + \textstyle \sum _{s = 1}^{\emph{S}} \op^{(b_s)} \EP_{m}^{(b_s)}},
\end{equation}
\begin{equation}\label{SINR-SC-3}
\textstyle \gamma_{s}^{(b_0)}(\mathbf{\Lambda}|\mathbf{\Theta}) \xrightarrow{a.s.} \textstyle \frac{\us_s^{(b_0)} p_{s}^{(b_0)}}{1 + \textstyle \sum _{s' = 1, s'\neq s}^{\emph{S}} \op^{(b_{s'})} \EP_{s}^{(b_{s'})}},
\end{equation}
\begin{equation}\label{SINR-SUE-3}
\textstyle \gamma_{{u}}^{(b_s)}(\mathbf{\Lambda}|\mathbf{\Theta}) \xrightarrow{a.s.} \textstyle \frac{ \op^{(b_s)} \us^{(b_{s})}_{u} \pc_u^{(b_s)} } {1 + \textstyle \sum _{s' = 1, s'\neq s}^{\emph{S}} \op^{(b_{s'})} \EP_{u}^{(b_{s'})}}.
\end{equation}
Moreover, we obtain the closed-form expression for the transmit power constraint, i.e.,
\begin{equation}
\displaystyle  \frac{1}{N} \textstyle \sum _{k = 1}^{{K}} \frac{ {p}_{k}^{(b_0)}}{  \Omega_{k}^{} } - P^{(b_0)} \leq 0. \notag
\end{equation}
\end{theorem}
Although the closed-form expressions of average data rate and transmit power are obtained, our problem considers a time-average optimization with a large number of control variables, and dynamic traffic load over the convex region for a given composite control variable $\mathbf{\Lambda}$ and $\mathbf{\Theta}$. Our aim is to maximize the aggregate network utility subject to queue stability in which the well-known Lyapunov optimization yields an utility throughput optimality and stability~\cite{neely2010S}. Hence, we apply the drift-plus-penalty technique~\cite{neely2010S} to solve load balancing, operation mode selection, and power allocation problems.
\section{ Lyapunov Optimization Framework}
\label{LOF}
The network operation is modeled as a queueing network that operates in discrete time $t \in \{0, 1, 2, \dots \}$. Let $a_k(t)$ denote the bursty data arrival destined for each user $k$, $\text{i.i.d}$ over time slot $t$.
Let
$\mathbf{Q}(t)$ denote the vector of transmission queue blacklogs at MBS at slot $t$. The queue evolution is given by
\begin{equation}\label{queueQ}
{Q}_{k}(t + 1) = \text{max}~[{Q}_{k}(t) - {r}_{k}(t), \quad 0] + {a}_{k}(t),\quad\forall~k \in \mathcal{K}.
\end{equation} 
Here, we consider the bound of the traffic arrival of user $k$ is bounded such that $0 \leq {a}_{k}(t) \leq a_{k}^{\text{max}}$, for some constant $a_{k}^{\text{max}} < \infty$. Futuremore, let $r_k^{\mathrm{max}}(t)$ be the
upper bound of data rate for user $k$ at time slot $t$, such that $r_k^{\text{max}}(t) \leq a_{k}^{\mathrm{max}}$. The set at constraint~(\ref{eq-02}) is replaced by an another equivalent set by introducing
$\text{auxiliary variables}$ $\boldsymbol{\varphi}(t) \in \mathcal{R}$, $\boldsymbol{\varphi}(t) = \big(\varphi_{1}(t), \ldots, \varphi_{K}(t) \big)$ that satisfies $\bar{\varphi}_{k} \leq \bar{r}_{k}$, where
$\bar{\varphi}_{k} \triangleq \lim_{t \to \infty} \frac{1}{t} \textstyle \sum_{\tau = 0} ^ {t-1} \mathbb{E} \big[ \varphi_{k}(\tau)  \big] $. The evolution of wireless backhaul queue is rewritten as
\begin{equation}\label{queueD}
{D}_{s}(t + 1) = \mathrm{max}~[{D}_{s}(t) + \varphi_{s+M}(t) - {r}_{\sue_s}^{(b_s)}(t), \quad 0], \quad\forall~s \in \mathcal{S}.
\end{equation}
For a given $\mathbf{\Lambda}$ and $\mathbf{\Theta}$, the optimization problem~(\ref{eq:Obj-Formulate-0}) subject to the network stability and dynamic backhaul can be posed as
\begin{subequations}\label{eq:Obj-Formulate-1}
\begin{align}
    \underset{\bar{\boldsymbol{\varphi}}}{\text{min}}
        & \quad - {f}_0 ( \bar{\boldsymbol{\varphi}}) \label{eq-1a} \\
    \text{subject to}
    &  \quad \bar{\varphi}_{k} - \bar{r}_{k} \leq 0,\quad\forall~k \in \mathcal{K},  \label{eq-1b} \\
    & \quad (\ref{UA-Condition}), \quad \bar{\mathbf{D}} < \infty, \mathbf{\bar{Q}} < \infty. \label{eq-1c}
\end{align}
\end{subequations}
In order to ensure the inequality constraint~(\ref{eq-1b}), we introduce a virtual queue vector ${Y}(t)$ which evolves as follows
\begin{equation}\label{queueY}
{Y}_{k}(t + 1) = \mathrm{max}~[{Y}_{k}(t) + \varphi_{k}(t) - {r}_{k}(t),\quad0], \quad\forall~k \in \mathcal{K}.
\end{equation}
We define the queue backlog vector as $\mathbf{\Sigma}(t) = \big[ \mathbf{Q}(t), \mathbf{Y}(t), \mathbf{D}(t)\big]$ (whereas the stability of $\mathbf{\Sigma}(t)$ yields all constraints of problem~(\ref{eq:Obj-Formulate-1})
are hold). The Lyapunov function can be written as
\begin{equation}
{L}(\mathbf{\Sigma}(t)) \triangleq \frac{1}{2} \big[ \textstyle \sum _{k = 1}^{\emph{K}} {Q}_{k}(t)^{2} + \textstyle \sum _{k = 1}^{\emph{K}} {Y}_{k}(t)^{2} + \textstyle \sum _{s = 1}^{\emph{S}} {D}_{s}(t)^{2}\big].\notag
\end{equation}
For each time slot $t$, $\mathbf{\Delta}(\mathbf{\Sigma}(t))$ denotes the Lyapunov drift, which is given by
\begin{equation} 
\mathbf{\Delta}(\mathbf{\Sigma}(t)) \triangleq \mathbb{E} \big [ {L}(\mathbf{\Sigma}(t+1))  - {L}(\mathbf{\Sigma}(t))|
\mathbf{\Sigma}(t) \big].\notag
\end{equation}

\begin{figure}
    \centering
    \includegraphics[scale=0.6]{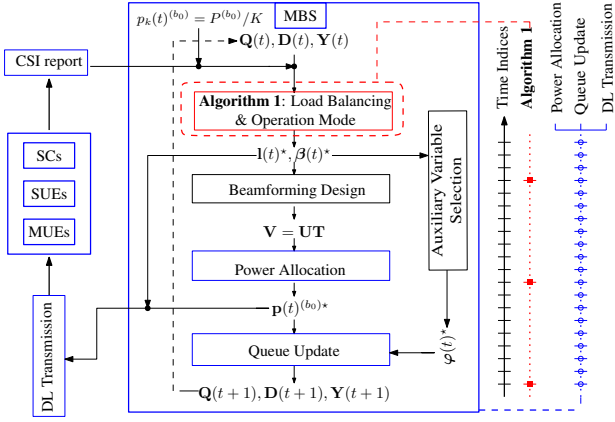}
    \caption{Joint load balancing and interference mitigation algorithm.}
    \label{Algorithm0}
\end{figure}

Noting that $\mathrm{max}[a,0]^2 \leq  a^2$ and $(a \pm b)^2 \leq a^2 \pm 2ab + b^2$ for any real positive number $a, b$, and thus, by neglecting the index $t$ we have:
\begin{equation}
(\mathrm{max}~[{Q}_{k} - r_{k}, \quad 0] + {a}_{k})^{2} - {Q}_{k}^2 \leq 2 {Q}_{k}(a_{k}- {r}_{k}) + (a_{k}- {r}_{k})^{2}, \notag
\end{equation}
\begin{equation}
\mathrm{max}~[{Y}_{k} + \varphi_{k}- {r}_{k}, \quad 0]^{2} - {Y}_{k}^2 \leq 2 {Y}_{k}(\varphi_{k}- {r}_{k}) + (\varphi_{k}- {r}_{k})^{2},\notag
\end{equation}
 \begin{align}\mathrm{max}~[{D}_{s} + \varphi_{s+M} &  - {r}_{\sue_{s}}^{(b_s)}(t), \quad 0]^{2} - {D}_{s}^2 \leq  2 {D}_{s}(\varphi_{s+M} \notag \\ & -  {r}_{\sue_{s}}^{(b_s)}(t) )  + (\varphi_{s+M}-   {r}_{\sue_{s}}^{(b_s)}(t)
)^{2}. \notag
\end{align}
We assume that $\boldsymbol{\varphi}_k \in \mathcal{R}$ and a feasible $\mathbf{\us}$ for all $t$ and all possible $\mathbf{\Sigma}(t)$, we have
\begin{align}\label{eq:Lya-Driff-Pen}
\mathbf{\Delta}(\mathbf{\Sigma}(t)) \leq & ~\Psi +  \textstyle \sum _{k = 1}^{\emph{K}} {Q}_{k}(t) \mathbb{E} \Big[  {a}_{k}(t) - {r}_{k}(t) | \mathbf{\Sigma}(t) \Big ] \notag
\\ &+ \textstyle \sum _{s = 1}^{\emph{S}} {D}_{s}(t)\mathbb{E} \big[  {\varphi}_{s+M}(t)  - {r}_{\sue_{s}}^{(b_s)}(t) |\mathbf{\Sigma}(t) \big ] \notag
\\ &+ \textstyle \sum _{k = 1}^{\emph{K}} {Y}_{k}(t) \mathbb{E} \big[ \varphi_{k}(t) - r_{k}(t) | \mathbf{\Sigma}(t) \big].
\end{align}
Here $\mathbf{\Delta}(\mathbf{\Sigma}(t)) \leq  \Pi$, where  $\Pi$ represents the $\text{R.H.S}$ of~(\ref{eq:Lya-Driff-Pen}), and $\Psi$ is a finite constant that satisfies $\Psi \geq      \frac{1}{2} \sum _{k =
1}^{\emph{K}}\mathbb{E} \big[  \big({a}_{k}(t) - {r}_{k}(t)\big)^2 | \mathbf{\Sigma}(t) \big ] +   \frac{1}{2} \sum _{k = 1}^{\emph{K}}  \mathbb{E} \big[  \big(\varphi_{k}(t) - {r}_{k}(t)\big)^2 | \mathbf{\Sigma}(t) \big ] +
\frac{1}{2} \sum _{s = 1}^{\emph{S}}\mathbb{E} \big[  \big( {\varphi}_{s+M}(t) - {r}_{s}^{\sue_{s}}(t) \big)^2 | \mathbf{\Sigma}(t)\big ]$, for all $t$ and all possible $\mathbf{\Sigma}(t)$.
We apply the Lyapunov drift-plus-penalty technique~\cite{neely2010S}, where the solution of~(\ref{eq:Obj-Formulate-1}) is obtained by minimizing the Lyapunov drift and a penalty from the objective function, i.e.,
\begin{equation} \mathrm{min}~\Pi-\nu \mathbb{E}[{f}_0
( \boldsymbol {\varphi}(t) )].\notag
\end{equation}
Here, the parameter $\nu$ is chosen as non-negative constant to control optimal minimization solution~\cite{neely2010S}. Since $\Psi$ is finite, the problem is to minimize the below expression
subject to the convex set hull, given by~(\ref{FinalEQ}).
\begin{floatEq}
\begin{equation}\label{FinalEQ}
 \Big[ \big[\overbrace{-\textstyle \sum _{k} \big ( {Q}_{k}(t) + {Y}_{k}(t) \big) {r}_{k}(\mathbf{\Lambda}(t)) }^{\text{Impact of network queue, virtual queue, and } \mathbf{\Lambda}} \big]_{1\star} \overbrace{-\textstyle
\sum _{s} {D}_{s}(t)  {r}_{\sue_{s}}^{(b_s)}(\op^{(b_s)}(t))}^{\text{Impact of SC queue and } \boldsymbol{\op}}\Big]_{2\star}
+ \Big[ \overbrace{\textstyle \sum _{k} {Y}_{k}(t) \varphi_{k}(t) +  \textstyle \sum _{s} {D}_{s}(t) {\varphi}_{s+M}(t)}^{\text{Impact of virtual queue, SC queue, and auxiliaries}}  \overbrace{- \nu {f}_0 ( \boldsymbol
{\varphi}(t) )}^{\text{penalty}}\Big]_{3\star}.
\end{equation}
\end{floatEq}
Note that~(\ref{FinalEQ}) is decoupled over user association, user scheduling, and operation mode variables ($2\star$), auxiliary variables ($3\star$), and precoder and power
allocation variables ($1\star$), respectively as in~(\ref{FinalEQ}). Hence, the respective variables can be found independently by minimizing the individual term at each time. Fig.~\ref{Algorithm0} summarizes the relationship
among various subproblems.

\subsection{Joint Load Balancing and Operation Mode Selection}
\label{loadbalancing}
First, the problem of joint load balancing and FD-enabled SC operation mode selection in ($2\star$) is cast as the minimization problem below.
\begin{subequations}\label{Original-US}
\begin{align}
    \underset{ \mathbf{l},\boldsymbol{\op}  }{\text{min}}
        &    - \textstyle \sum _{k = 1}^{\emph{K}} A_{k}(t) \log \big( 1 +  \us_k^{(b_0)}(t) \frac{ p_{k}^{(b_0) } (1 - \tau_k^2)} {1 + \textstyle \sum _{s = 1}^{\emph{S}} \op^{(b_s)} \EP_{k}^{(b_s)} } \big) \nonumber\\
        & - \textstyle \sum _{s = 1}^{\emph{S}} {D}_{s}(t)  \textstyle  \log  \big( 1 +  \op^{(b_s)}(t) \frac{\us_{c_s}^{(b_s)}(t) p_{c_s}^{(b_s)} }{1 + \textstyle \sum _{s'\neq s}^{\emph{S}} \op^{(b_{s'})} \EP_{c_{s'}}^{(b_{s'})}  } \big) \label{eq-OUS1}\\
    \text{subject to}
    &   \quad \us_j^{(b_{s})}(t)\in \{0, 1\}, \forall~j \in \mathcal{K} \cup \mathcal{C},\quad\forall~b_s \in \mathcal{B},\label{eq-OUS2}\\
    &   \quad \op^{(b_s)}(t)\in \{0, 1\},  \emph{N}_{s}^{\mathrm{tx}}(t) \leq \emph{N}_{s}^{\mathrm{au}}, \forall~s \in \mathcal{S},\label{eq-OUS3}\\
    &   \quad \textstyle \sum _{s = 0}^{\emph{S}} \us_{m}^{(b_{s})}(t)  \leq 1, \forall~m \in \mathcal{M}, \notag \\
    &   \quad \textstyle \sum _{m = 1}^{\emph{M}} \us_{m}^{(b_{s})}(t)   + \us_{c_s}^{(b_{s})}(t)  =  \emph{N}_{s}^{\mathrm{tx}}(t), \label{eq-OUS4}\\
    &   \quad (\ref{UA-Condition}), r_k(t) \in \mathcal{R},  \notag \\
    &   \quad \textstyle \sum_{k = 1}^{\emph{K}} \us_{k}^{(b_0)}(t) + \sum_{s = 1}^{\emph{S}} \emph{N}_{s}^{\mathrm{tx}}(t) \leq \emph{N},\label{eq-OUS6}\\
    &   \quad \textstyle \sum_{k = 1}^{\emph{K}} \sum_{s =1}^{\emph{S}} \us_k^{(b_0)}(t)  \op^{(b_s)}(t)  \EP_{k}^{(b_s)}(t)  \leq  \epsilon_{o}, \label{eq-OUS7}
\end{align}
\end{subequations}
where $A_{k}(t) = {Q}_{k}(t) + {Y}_{k}(t)$. This problem is a non-convex program with binary variables. It turns out this problem has a hidden convexity structure and the non-convex terms can be iteratively approximated by its convex upper bound via an iterative SCA method. The motivations of utilizing the SCA method are due to (i) its low complexity and fast convergence~\cite[Lemma 3.5]{beck2010seq} and (ii) the obtained solution which yields many relaxed variables are close to zero or one ~\cite{tran2012}. In this regard, we convexify this problem to find a sub optimal solution. First, we relax the binary constraints~(\ref{eq-OUS2}) and~(\ref{eq-OUS3}) to linear constraints as continuous variables. Secondly, at each iteration $i$ the non-convex constraint~(\ref{eq-OUS7}) is approximated by upper convex approximation, i.e., 
\begin{equation}\sum_{k=1}^{K} \sum_{s =1}^{\emph{S}} \big( \frac{\lambda_{ks}^{(i)} (\us^{(b_0)}_{k}(t))^{2} }{2} +
\frac{(\op^{(b_s)})^{2}(t)}{2 \lambda_{ks}^{(i)}} \big) \EP_{k}^{(b_s)}(t) - \epsilon_{o} \leq 0,\notag
\end{equation}
for every fixed positive value $\lambda_{ks}^{(i)}$. Finally, instead of minimizing the non-convex objective function
$(\ref{eq-OUS1})$ we convert it into a convex function by the followings. We minimize its upper bound by replacing the denominators, i.e., $1 + \textstyle \sum _{s = 1}^{\emph{S}} \op^{(b_s)} \EP_{m}^{(b_s)}$ with
largest bound, i.e., $1 + \epsilon_{0}$. Due to interference constraint~(\ref{eq-OUS7}), we obtain the upper bound as below
\begin{align}\label{non-obj}
&-  \sum _{k = 1}^{\emph{K}} A_{k}(t) \log \big( 1 +  \frac{\us_k^{(b_0)}(t) p_{k}^{(b_0) } (1 - \tau_k^2)} {1 + \epsilon_{0} } \big) \notag \\
&-   \sum _{s = 1}^{\emph{S}} {D}_{s}(t)  \log  \big( 1 +   \op^{(b_s)}(t) \frac{\us_{{c_s}}^{(b_s)}(t) p_{{c_s}}^{(b_s)} }{1 + \epsilon_{0}  } \big).\nonumber
\end{align}
Using the similar approach as convexifying the interference constraint~(\ref{eq-OUS7}), we convexify the second part of these objective function which still remains non-convex. We denote the lower bound of SINR of UE served SC $b_s$ as $\underline{\gamma}^{b_s} (t)$, let us set $\tilde{\us}_{{c_s}}^{(b_s)}(t) \triangleq \frac{{\us}_{{c_s}}^{(b_s)}(t) p_{u}^{(b_s)} }{1 + \epsilon_{0} }$. Then we have:
\begin{equation}\label{non-obj1}
\underline{\gamma}^{b_s} (t) \leq \op^{(b_s)}(t) \tilde{\us}_{{c_s}}^{(b_s)}(t),\quad\forall~s \in \mathcal{S},
\end{equation}
by introducing the new slack variable $\iota^{2}_{s}(t)$,~(\ref{non-obj1}) is equivalent to:
\begin{align}
 \frac{1}{4} \big ( \op^{(b_s)}(t) - \tilde{\us}_{{c_s}}^{(b_s)}(t) \big ) ^2 + \iota^{2}_{s}(t)  &\leq \frac{1}{4} \big ( \op^{(b_s)}(t) + \tilde{\us}_{{c_s}}^{(b_s)}(t) \big ) ^2, \label{eq-socp1}\\
\text{and} \quad \underline{\gamma}^{b_s} (t) & \leq \iota^{2}_{s}(t),\quad\forall~s \in \mathcal{S}. \label{eq-socp2}
\end{align}
where the constraint~(\ref{eq-socp1}) holds a form of the second-order cone inequalities (SOC), while the RHS of the set of constraints in~(\ref{eq-socp2}) are still non-convex, which can be approximated by using the iterative SCA method~\cite{beck2010seq}. We rewrite the constraint~(\ref{eq-socp2}) as
\begin{align}\label{eq-socp3}
\underline{\gamma}^{b_s} (t) \leq  \hat{\iota}^{(i)2}_{s}(t) + 2 \hat{{\iota}}_{s}^{(i)}(t) ({\iota}_{s}(t) -\hat{{\iota}}_{s}^{(i)}(t)),\quad\forall~s \in \mathcal{S},
\end{align}
where at iteration $i+1$, we update $\hat{\iota}^{(i+1)}_{s}(t)$ such that $\hat{\iota}^{(i+1)}_{s}(t) = {\iota}^{(i)}_{s}(t)$.
Hence, the optimal value of $\mathbf{\Lambda}^{o}$ is given by
\begin{subequations}\label{Optimal-US}
\begin{align}
    \underset{\mathbf{l},\boldsymbol{\op}}{\text{min}}
        & \quad - \textstyle \sum _{k = 1}^{\emph{K}} A_{k}(t) \log \big( 1 +  \us_k^{(b_0)}(t) \frac{ p_{k}^{(b_0) } (1 - \tau_k^2)} {1 + \epsilon_{0} } \big) \notag \\
        & \quad - \textstyle \sum _{s = 1}^{\emph{S}} {D}_{s}(t)  \log \big( 1 + \underline{\gamma}^{b_s} (t)\big)
\label{eq-us1}\\
    \text{subject to}
        & \quad   \us_j^{(b_{s})}(t)\in [0, 1], \forall~j \in \mathcal{K} \cup \mathcal{C}, \forall~b_s \in \mathcal{B},\label{eq-us2}\\
        & \quad  \op^{(b_s)}(t)\in [0, 1],  \emph{N}_{s}^{\mathrm{tx}}(t) \leq \emph{N}_{s}^{\mathrm{au}},  \forall~s \in \mathcal{S},\label{eq-us3}\\
        & \quad (\ref{eq-OUS4}), (\ref{eq-OUS6}),(\ref{eq-socp1}), (\ref{eq-socp3}), \label{eq-us4}\\
        & \quad  \textstyle \sum_{k = 1}^{\emph{K}} \textstyle \sum_{s =1}^{\emph{S}} \big( \frac{\lambda_{ks}^{(i)} (\us^{(b_0)}_{k}(t))^{2} }{2} \notag\\
        & \quad + \frac{(\op^{(b_s)})^{2}(t)}{2 \lambda_{ks}^{(i)}} \big) \EP_{k}^{(b_s)}(t) - \epsilon_{o} \leq 0. \label{eq-us7}
\end{align}
\end{subequations}
At each time slot $t$, the approximated problem~\eqref{Optimal-US} is iteratively solved as in Algorithm~\ref{algLB}. We numerically observe that the SCA-based Algorithm~\ref{algLB}
converges quickly within few iterations and yields a continuous relaxation solution of many user association and operation mode variables close or equal to binary. To ensure that all users will be served, when performing Algorithm~\ref{algLB} each user is
assumed to receive the same transmit power to find the best scheduled users. Moreover, the scheduling will be performed in a long-term period, while the power allocation problem is executed in a short-term period.
\begin{algorithm}                   
\caption{Joint load balancing and operation mode algorithm}
\label{algLB}
\begin{algorithmic}                    
    \STATE Initialization $i := 0$, $\lambda_{ks}^{(i)}, \hat{\iota}^{(i)}_{s} := \text{randomly positive that satisfy all constraints}$.
    \REPEAT
        \STATE $\text{Solve}$~(\ref{Optimal-US}) with $\lambda_{ks}^{(i)}, \hat{\iota}^{(i)}_{s}$ to get optimal value $\mathbf{\Lambda}^{o\star} = \{ \mathbf{\us}^{\star}, \boldsymbol{\op}^{\star} \}$.
        \STATE $\text{Update}$ $\mathbf{\Lambda}^{o(i)} := \mathbf{\Lambda}^{o\star}$ and $\lambda_{ks}^{(i+1)} := \frac{\op^{(b_s)(i)}} {\us^{(b_0)(i)}_{k}}$; $\hat{\iota}^{(i+1)}_{s} := {\iota}^{(i)}_{s}$; $i:= i + 1$.
    \UNTIL{\text{Convergence}}
\end{algorithmic}
\end{algorithm}
Since the objective function of the problem~(\ref{Optimal-US}) is a maximum weighted matching problem with respect to linear or square
function, we use a low-complexity binary search algorithm~\cite{li2014e} to obtain the final solutions with lower dimensions. Let $\emph{K}_{1} = \{j,s | \us_j^{(b_{s})\star}, \op^{(b_s)\star} = 1\}$, $\emph{K}_{\text{uct}} =\{ j, s | \xi \leq \us_j^{(b_{s})\star}, \op^{(b_s)\star} \leq 1\}$, and $\emph{K}_{0} =\{ j, s | \us_j^{(b_{s})\star}, \op^{(b_s)\star} \leq  \xi\}$ denote set of selected variables, set of uncertain variables, set of removed variables, respectively, where $\xi$  is some small threshold.  First, we determine the set $\emph{K}_{1}$, $\emph{K}_{\text{uct}}$, and $\emph{K}_{0}$ based on $\xi$. Then, we consider to select among the uncertain variables in $\emph{K}_{\text{uct}}$. By sorting $\emph{K}_{\text{uct}}$ in a descending order, a loop starts by selecting one by one variable based on their largest weights according to the objective function.  We set the value uncertain variable to 1, and add it to $\emph{K}_{1}$, if it satisfies the antennas (\ref{eq-us4}) and interference (\ref{eq-us7}) constraints. If it does not satisfy the constrains, we add it to $\emph{K}_{0}$. The loop stops until reaching the last uncertain variable or the antennas constrain is over. Finally, $\emph{K}_{1}$ is kept, while $\emph{K}_{0}$ and $\emph{K}_{\text{uct}}$ are removed.
\subsection{The Selection of Auxiliary Variable}
\label{AV}
The optimal auxiliary variable from ($3\star$) is computed by
\begin{subequations}\label{eq:Optimal-AR}
\begin{align}
    \underset{\boldsymbol{\varphi}(t)}{\text{min}}
        &  \textstyle \sum _{k = 1}^{\emph{K}} {Y}_{k}(t) \varphi_{k}(t) + \textstyle \sum _{s = 1}^{\emph{S}} {D}_{s}(t) {\varphi}_{s+M}(t) \\
        &  \quad -  \nu  \textstyle \sum _{k = 1}^{\emph{K}} \omega_{k} (t) f( \varphi_{k}(t) ) \\
    \text{subject to}
        & \quad \varphi_{k}(t) \leq a_k^{\mathrm{max}}(t).
\end{align}
\end{subequations}
Since the above optimization problem is convex, let $\varphi_{k}^{\ast}(t)$ be the optimal solution obtained by the first order derivative of the objective function of (\ref{eq:Optimal-AR}). With a logarithmic utility
function, we have:
\[ \varphi_{k}^{\ast}(t) =  \begin{cases} \frac{\nu \omega_k(t)}{{Y}_{k}(t)} & \quad \text{if } k \leq \emph{M},\\
 \frac{\nu \omega_k(t)}{{Y}_{k}(t) + {D}_{k-\emph{M}}(t)} & \quad \text{otherwise.} \end{cases}\]

The optimal auxiliary variable is $\ \min \{\varphi_{k}^{\ast}(t), a_k^{\mathrm{max}}(t)\} $. 
\subsection{Interference Mitigation and Power Allocation}
\label{IM-WB}
For given scheduled users, the precoder $\mathbf{U}$ is found by solving~(\ref{zero-ICI}). Finally, problem~(\ref{eq:Obj-Formulate-1}) is decomposed to find the transmit power $p_k^{(b_0)}(t)$ from ($1\star$) that minimizes:
\begin{subequations}\label{WB0}
\begin{align}
    \underset{\mathbf{p}(t)}{\text{min}}
        & \quad - \textstyle \sum _{k = 1}^{\emph{K}} {A}_{k}(t)  r_{k}(\mathbf{p}(t)) \\
    \text{subject to}
        & \quad  \frac{1}{\emph{N}} \textstyle \sum _{k = 1}^{\emph{K}} \frac{ {p}_{k}^{(b_0)}(t) }{ \Omega_{k}(t) } - P^{(b_0)} \leq 0, \notag \\
        & \quad {p}_{k}^{(b_0)}(t) \geq 0, \forall~k \in \mathcal{K}. \notag
\end{align}
\end{subequations}
The objective function~(\ref{WB0}) is rewritten as $n(\mathbf{p}(t)) = -  \textstyle \sum _{k = 1}^{\emph{K}} A_{k}(t) \log \big( 1 +   p_{k}^{(b_0)}(t) n_k(t)  \big)$, where $n_k(t) = \frac{\us_k^{(b_0)}(t) (1 - \tau_k^2)}
{1 +  \textstyle \sum _{s = 1}^{\emph{S}} \op^{(b_s)}(t) \EP_{k}^{(b_s)} (t)}$. The objective function is strictly convex for $ {p}_{k} ^ {(b_0)}(t) \geq 0, \forall k \in \mathcal{K}$, and the constraints are compact. Hence,
the optimal solution of $\mathbf{p}^{\star}(t)$ exists, the Lagrangian function is written as $\mathcal{L}(\mathbf{p}(t), \mu_{0}) = n(\mathbf{p}(t)) + \mu_{0} \mathbf{g}(\mathbf{p}(t))$, where $\mu_0 \geq 0$ is the KKT
multiplier. The KKT conditions are
\begin{equation}\label{KKT1}
\nabla  n(\mathbf{p}(t))^{T} + \mu_{0} \textstyle  \frac{1}{\emph{N}} \textstyle \sum _{k = 1}^{\emph{K}} \frac{1} { \Omega_{k}(t) }  = 0.
\end{equation}
\begin{equation}
\mu_{0} \Big(\textstyle  \frac{1}{\emph{N}} \textstyle \sum _{k = 1}^{\emph{K}} \frac{ {p}_{k}^{(b_0)}(t)} { \Omega_{k}(t) } - P^{(b_0)} \Big) = 0. \label{KKT2}
\end{equation}
\begin{equation}
 \frac{1}{\emph{N}} \textstyle \sum _{k = 1}^{\emph{K}} \frac{ {p}_{k}^{(b_0)}(t) }{ \Omega_{k}(t) } - P^{(b_0)} \leq 0, \qquad
- \mathbf{p}(t) \leq 0, \mu_{0} \geq 0. \label{KKt4}
\end{equation}
Here, $\nabla  n(\mathbf{p}(t))^{T} = (n'({p}_{1}^{(b_0)}(t)), \ldots, n'({p}_{\emph{K}}^{(b_0)}(t)))$ where $n'({p}_{k}^{(b_0)}(t))=  \frac{-A_{k}(t)  n_k(t)}{1 + p_k^{(b_0)}(t)  n_k(t)}$. For $\mu_{0} \neq 0$, from
(\ref{KKT1}), obtaining
\begin{equation}\label{KKTS1}
p_k^{(b_0)}(t) = \max [\frac{A_k \emph{N} \Omega_{k}(t) }{ \mu_{0}} - \frac{1}{n_k(t)}, 0],
\end{equation}
from (\ref{KKT2}) and (\ref{KKTS1}) we derive $\mu_{0}$. Finally, the optimal value of $p_k(t)^{(b_0)\star}$ is obtained with (\ref{KKTS1}).
\subsection{Queue Update}
\label{QU}
Update the virtual queues ${Y}_k(t)$ and ${D}_s(t)$ according to~(\ref{queueY}) and (\ref{queueD}), and the actual queue ${Q}_k(t)$ in~(\ref{queueQ}).

$\textbf{Theorem~\ref{Theo1}}$ is provided to show the performance analysis of network utility maximization based on Lyapunov framework and prove that the queues are stable.
\begin{theorem}\label{Theo1}[Optimality] Assume that all queues are initially empty. For arbitrary arrival rates, the operation mode and load balancing is chosen to satisfy~(\ref{FinalEQ}) and the rate regime. For a given
constant $\chi \geq 0$, the network utility maximization with any $\nu > 0$ provides the following utility performance with $\chi-\textit{approximation}$
\[
{f}_0 \geq {f}^{\ast}_{0} - \frac{ \Psi + \chi}{\nu},
\]
where ${f}^{\star}_{0}$ is the optimal network utility over the rate regime. While the strong stability of the virtual queues and the network queues is given by
\begin{equation}\label{Bound1}
{Q}_{k}(t) \leq \nu \omega_k(t) \pi_k + 2 a_{k}^{\mathrm{max}}, \quad \forall   t \geq 0, \quad \forall k \in \mathcal{K}, \notag
\end{equation}
\begin{equation}\label{Bound2}
{Y}_{k}(t)  \leq  \nu \omega_k(t) \pi_k + a_{k}^{\mathrm{max}}, \quad \forall   t \geq 0, \quad \forall k \in \mathcal{K}, \notag
\end{equation}
\begin{equation}\label{Bound3}
{D}_{s}(t)  \leq  \nu \omega_{s+M}(t) \pi_{s+M} + a_{s+M}^{\mathrm{max}}, \quad \forall   t \geq 0, \quad \forall s \in \mathcal{K}. \notag
\end{equation}
\end{theorem}
$\textit{Proof:}$ Proof can be found in~\cite{neely2010S} and is omitted for the sake of brevity.

\subsection{~Relaxation of Utility Function}
\label{Relaxation}
Note that the previous discussion explains how to transform the above non-convex program \eqref{Original-US} as a generic convex program. Although it can be solved by using the modern solvers, generally it requires more
computation time. In order to reduce the computation time and speed up the optimization convergence, we relax the log function of the objective function~(\ref{eq-us1}) by a set of linear functions.
Moreover, in order to model and solve the problem efficiently, we use YALMIP toolbox~\cite{yalmip2004}, which can employ SDPT3~\cite{sdpt31999} or MOSEK~\cite{mosek2015} as internal solver. In general, we rewrite the log function as
\begin{equation}\label{LogApp}
1 + \gamma(l_{k}) \geq e^{r_k}, \notag
\end{equation}
where $\gamma(l_{k})$ is the $\mathrm{SINR}$ as a function of $l_k$. By using the results of approximation of second order cone programming~\cite{ben2001SOC}, \cite{nguyen2015}, (\ref{LogApp}) can be approximated by a set of following linear equations
\begin{align}\label{LinearEqs}
&\quad 1 + \gamma(l_{k}) \geq \kappa_{0}, \quad 1 + \kappa_{1}    \geq \|[1 - \kappa_{1} ~~~ 2 + r_k/2^{i-1}]\|_2, \nonumber \\
&\quad 1 + \kappa_{2}    \geq \|[1 - \kappa_{2} ~~~ 5/3 + r_k/2^{i}]\|_2,  \nonumber\\
&\quad 1 + \kappa_{3}    \geq \|[1 - \kappa_{3} ~~~ 2\kappa_{1}]\|_2, \nonumber\\
&\quad    \kappa_{4}    \geq \kappa_{2}  + \kappa_{3}/24 + 19/72, \nonumber\\
& \quad 1 + \kappa_{j}    \geq \|[1 - \kappa_{j} ~~~ 2\kappa_{j-1}]\|_2,\quad j = \{5, \cdots, i+3\},\nonumber\\
&\quad 1 + \kappa_{0}    \geq \|[1 - \kappa_{0} ~~~ 2\kappa_{i+3}]\|_2, \nonumber
\end{align}
where $\{\kappa_j\}_{j = 0, 1, \cdots, i+3}$, are new introduced variables, and the accuracy of the approximation depends on $i$. We numerically observe that the error accuracy is less than $10^{-5}$ when $i=10$.
\section{~~~~Numerical results}
\label{Evaluation}
In this section Monte Carlo simulations are carried out in order to evaluate the system performance of our proposed algorithm. To solve $\textbf{Algorithm~\ref{algLB}}$, we use YALMIP toolbox~\cite{yalmip2004} to model the optimization problem with SDPT3~\cite{sdpt31999} or MOSEK~\cite{mosek2015} as internal solver. For simulation, we consider the proportional fairness utility function, i.e., $f(\bar{r}_{k} ) = \log{(10^{-4} + \bar{r}_{k})}$~\cite{mo2000fair}. We denote our proposed user association algorithms for HetNet (resp. Homogeneous network) as $\textbf{HetNet-Hybrid}$ (resp. $\textbf{HomNet}$~\cite{wagner2012l}). Here,
$\textbf{HomNet}$~\cite{wagner2012l} refers to when the MBS serves both MUEs and SUEs without SCs. We compare our proposed algorithm with $\textbf{HomNet}$~\cite{wagner2012l} and with the previous work~\cite{Vu2016} ($\textbf{HetNet-Closed Access}$~\cite{Vu2016}). $\textbf{HetNet-Closed Access}$~\cite{Vu2016} case considers only joint
in-band scheduling and interference mitigation algorithm with fixed user association scheme (SCs are configured in closed subscriber group). The network performance are evaluated under the impact of the number of SCs per $\text{km}^{2}$, the number of MBS antennas $\emph{N}$, and the MBS transmit power levels $P^{(b_0)}$ at low and high frequency bands. We provide the convergence behaviour of the proposed method and validation of the approximation method.

\subsection{Simulation Environments} 
\label{Simulation}
Consider a HetNet scenario, where a MBS is located at the center of a square area, MUEs are randomly deployed within the coverage of the MBS (the minimum MBS-MUE distance is $35$ m). The SCs are uniformly distributed and one SUE per each SC is considered. The number of antennas at SCs $\emph{N}_{s}$ is greater than two, while we assume each SC can serve up to $\emph{N}_{s}^{\mathrm{au}} = 2$ UEs (including its own SUE). The path loss is modeled as a distance-based path loss with line-of-sight (LOS) model for urban environments~\cite{mW2014}. We first assume that the probability of obtaining LOS is very high to make the performance evaluation, while the effect of other channel models is studied later. The FD interference threshold $\epsilon_{o}$ is set to $5 \times 10^{-3}$ and the RZF parameter is $\alpha = 10^{-2}$. The data arrivals follow the Poisson distribution with the mean rate of $1$ Gbps, $100$ Mbps, and $20$ Mbps for $28$ GHz, $10$ GHz, and $2.4$ GHz, respectively. The parameter settings are summarized in~Table~\ref{parameter}.

\begin{table}[t]
\caption{Parameter settings} 
\centering
\begin{tabular}{ |l|l|l| }
\hline 
Path loss model~\cite{mW2014} & Values in dB & Bandwidth (B)\\
\hline 
LOS @ 28 GHz         & $61.4 + 20 \log(d)$     & $1$ GHz\\
LOS @ 10 GHz         & $55.25 + 18.5 \log(d)$     & $100$ MHz\\
LOS @ 2.4 GHz        & $17 + 37.6 \log(d)$     & $20$ MHz\\
\hline
\multicolumn{2}{ |c| }{Parameter} & Values\\ 
\hline 
\multicolumn{2}{ |c| }{ Maximum transmit power of MBS $P^{(b_0)}$} & $41$~dBm \\
\multicolumn{2}{ |c| }{ Channel quality  $\tau$}                   & $0.1$      \\
\multicolumn{2}{ |c| }{ SC antenna gain }                          & $5$ dBi \\
\multicolumn{2}{ |c| }{ Lyapunov parameter  $\nu$}                 & $2\times10^6$  \\
\hline
\end{tabular}
\label{parameter}
\end{table}
\subsection{~~~~Ultra-Dense Small Cells Environment}
\label{SubEvaluation1}
\begin{figure*}[!t]
	\centering
		\begin{minipage}{3in}
			\centering
			\includegraphics[width=0.95\textwidth,scale=0.5]{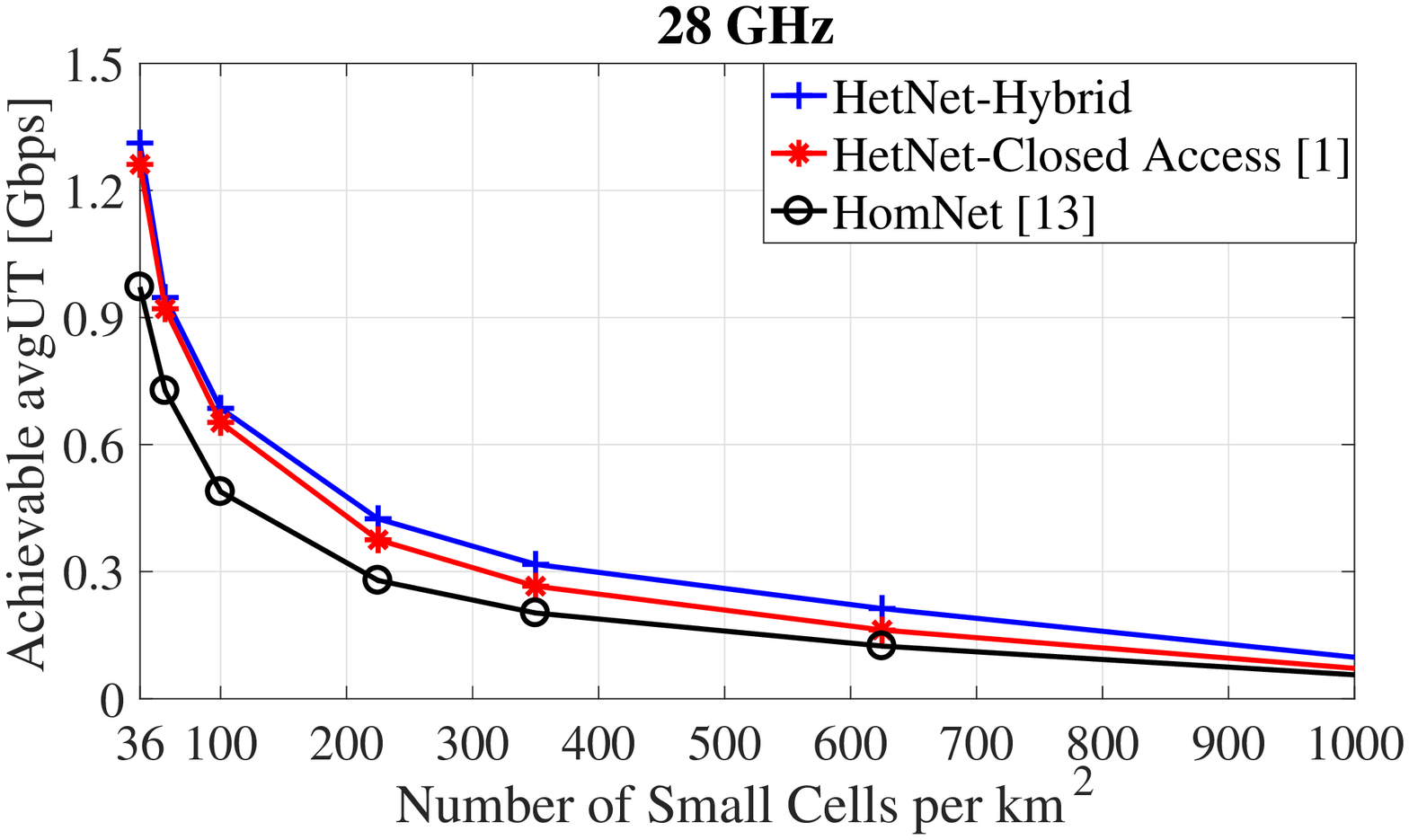} 
			\caption{Achievable avgUT versus number of small cells per $\text{km}^{2}$, $\emph{S}$, when scaling $\emph{K} = 2.5 \times \emph{S}$, $\emph{N} = 2 \times \emph{K}$.}\label{AvgUEDensity}
		\end{minipage}	
        \quad
	       \begin{minipage}{3in}
		  \centering
		  \includegraphics[width=0.95\textwidth,scale=0.5]{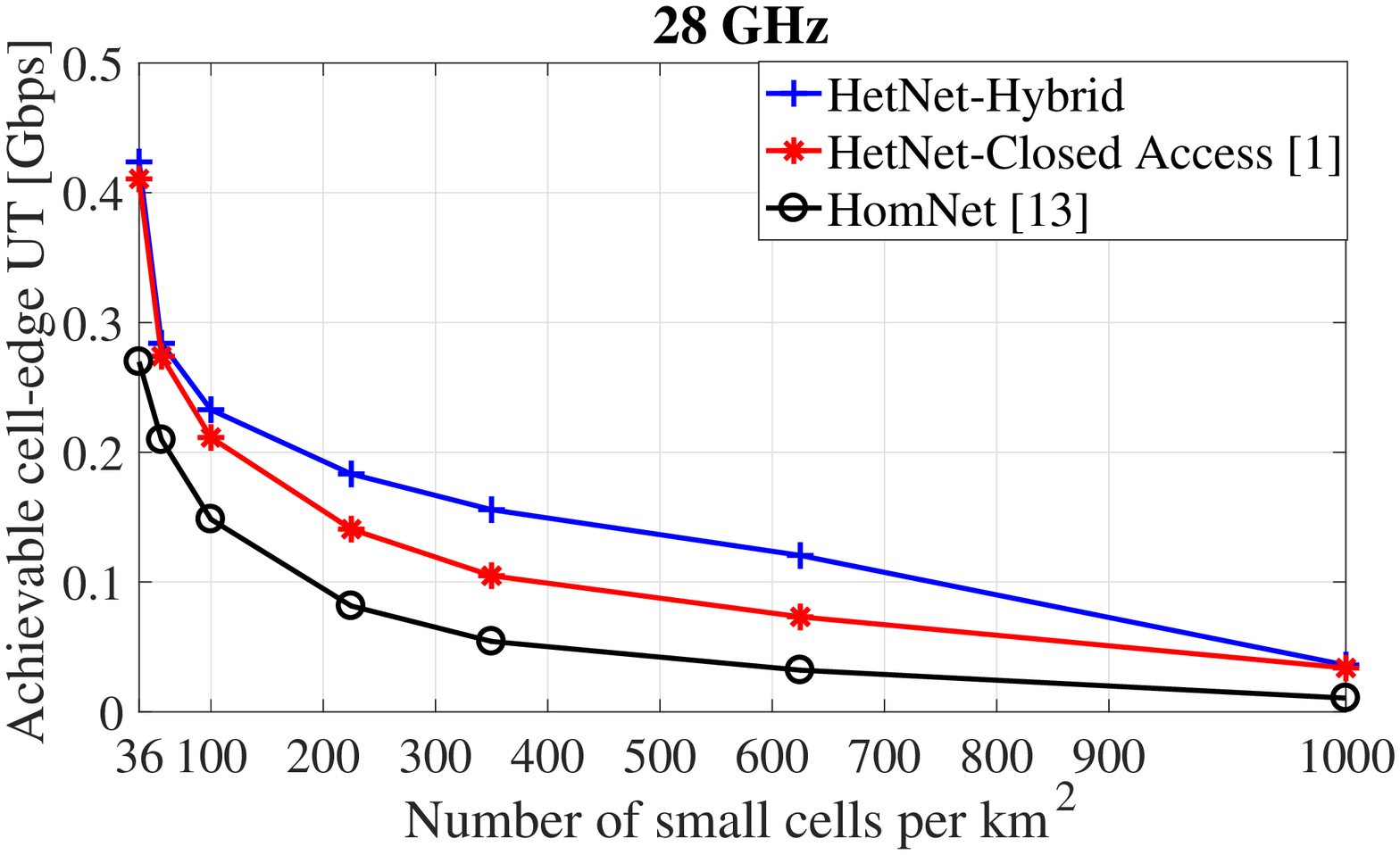} 
		  \caption{Achievable cell-edge UT versus number of small cells per $\text{km}^{2}$, $\emph{S}$, when scaling $\emph{K} = 2.5 \times \emph{S}$, $\emph{N} = 2 \times \emph{K}$.}\label{TotalUtiDensity}
	   \end{minipage}
\end{figure*}

To show the impact of network density, the average UE throughput (avgUT) and the cell-edge UE throughput (cell-edge UT) as a function of the number of SCs are shown in Fig.~\ref{AvgUEDensity} and Fig.~\ref{TotalUtiDensity}, respectively. The maximum transmit power of MBS and SCs is set to $41$ dBm and $32$ dBm, respectively\footnote{We reduce the maximum transmit power of MBS as compared to our previous work~\cite{Vu2016}, which used that of $43$ dBm. Hence, the average UT in this scenario is lower than~\cite{Vu2016}.}. In Fig.~\ref{AvgUEDensity} and Fig.~\ref{TotalUtiDensity}, the simulation is carried out in the asymptotic regime where the number of BS antennas and the network size (MUEs and SCs) grow large with a fixed ratio~\cite{S2014if}. In particular, the number of SCs and the number of SUEs are both increased from $36$ to $1000$ per $\text{km}^{2}$, while the number of MUEs is scaled up with the number of SCs, such that $\emph{M} = 1.5 \times \emph{S}$. Moreover, the number of transmit antennas at MBS and SCs  is set to $\emph{N} = 2 \times \emph{K}$ and $\emph{N}_{s} = 6$, respectively. We recall that when adding SCs we also add one SUE per one SC that increases the network load. Here, the total number of users is increased while the maximum transmit power is fixed, and thus, the per-user transmit power is reduced with $1/K$, which reduces the per-UE throughput. Even though the number of MBS antennas is increased with $K$, as shown in Fig.~\ref{AvgUEvsAntennas} and Fig.~\ref{Cell-Edge-UEvsAntennas}, the performance of massive MIMO reaches the limit as the number of antennas goes to infinity. It can be seen that with increasing network load, our proposed algorithm $\textbf{HetNet-Hybrid}$ outperforms baselines (with respect to the avgUT and the cell-edge UT) and the performance gap of the cell-edge UT is largest ($5.6\times$) when the number of SC per $\text{km}^{2}$ is 350, and is small when the number of SC per $\text{km}^{2}$ is too small or too large. The reason is that when the number of SCs per $\text{km}^{2}$ is too small, the probability for an MUE to find a open access nearby-SC to connect is low. With increasing the number of SCs per $\text{km}^{2}$ MUEs are more likely to connect with open access nearby-SCs to increase the cell-edge UT. However, when the number of SCs per $\text{km}^{2}$ is too large, the cell-edge UT performance of $\textbf{HetNet-Hybrid}$ is close to that of $\textbf{HetNet-Closed Access}$~\cite{Vu2016} due to the increased FD interference. Moreover, Fig.~\ref{AvgUEDensity} and Fig.~\ref{TotalUtiDensity} show that the combination of Massive MIMO and FD-enabled SCs improves the network performance; for instance, $\textbf{HetNet-Hybrid}$ and $\textbf{HetNet-Closed Access}$~\cite{Vu2016} outperform $\textbf{HomNet}$~\cite{wagner2012l} in terms of both the avgUT and the cell-edge UT. Our results provide good insight for network deployment: for a given target UE throughput, what is the optimal number of UEs to schedule and what is the optimal/maximum number of SCs to be deployed?
\subsection{~~~~Wireless Backhaul Impact versus Number of MBS Antennas} 
\label{SubEvaluation2}
\begin{figure*}[!t]
	\centering
	\begin{minipage}{3in}
		\centering
		\includegraphics[width=0.95\textwidth]{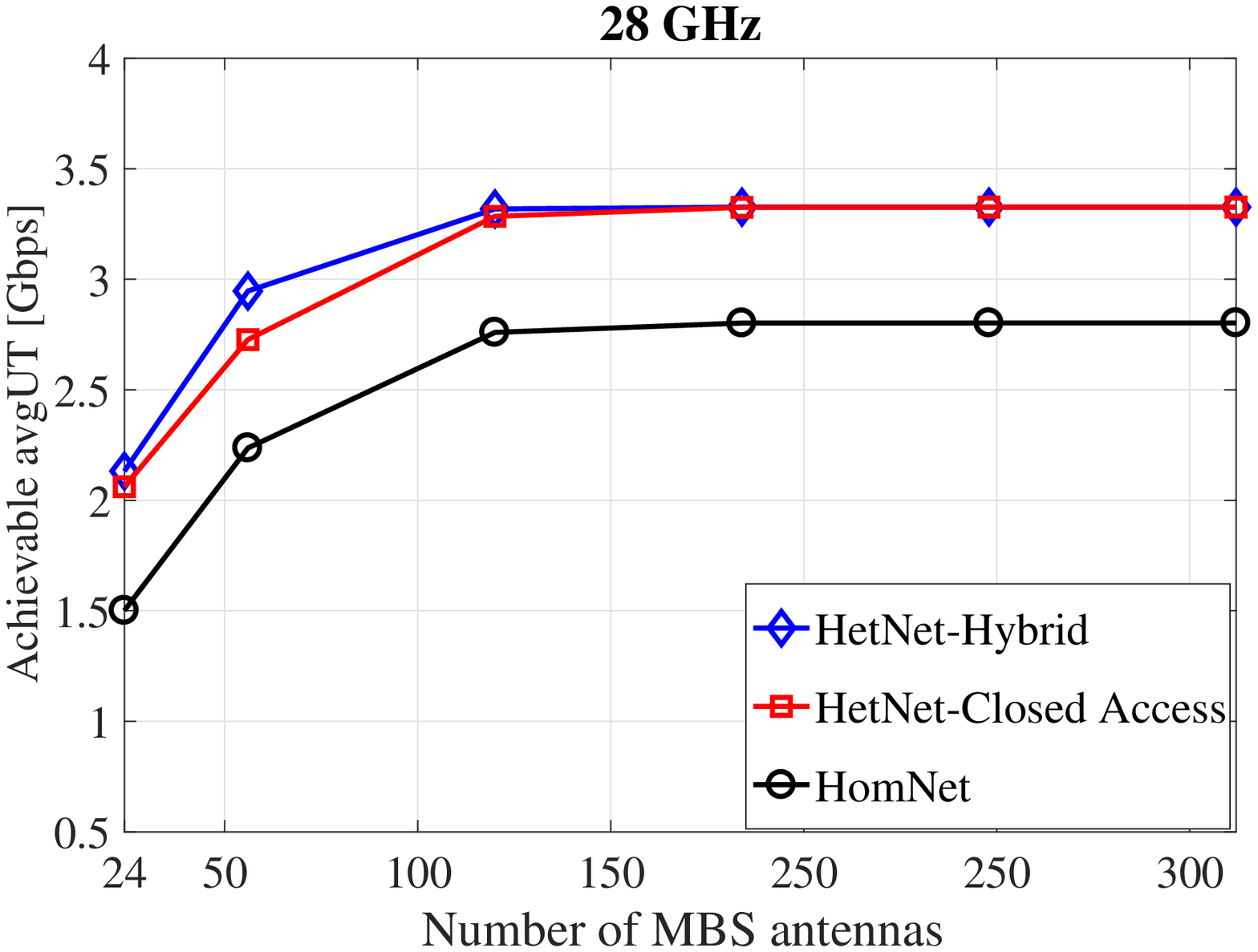} 
		\caption{Achievable avgUT versus $\emph{N}$, when $\emph{K} = 12$.} \label{AvgUEvsAntennas}
	\end{minipage}
	\begin{minipage}{3in}
		\centering
		\includegraphics[width=0.95\textwidth]{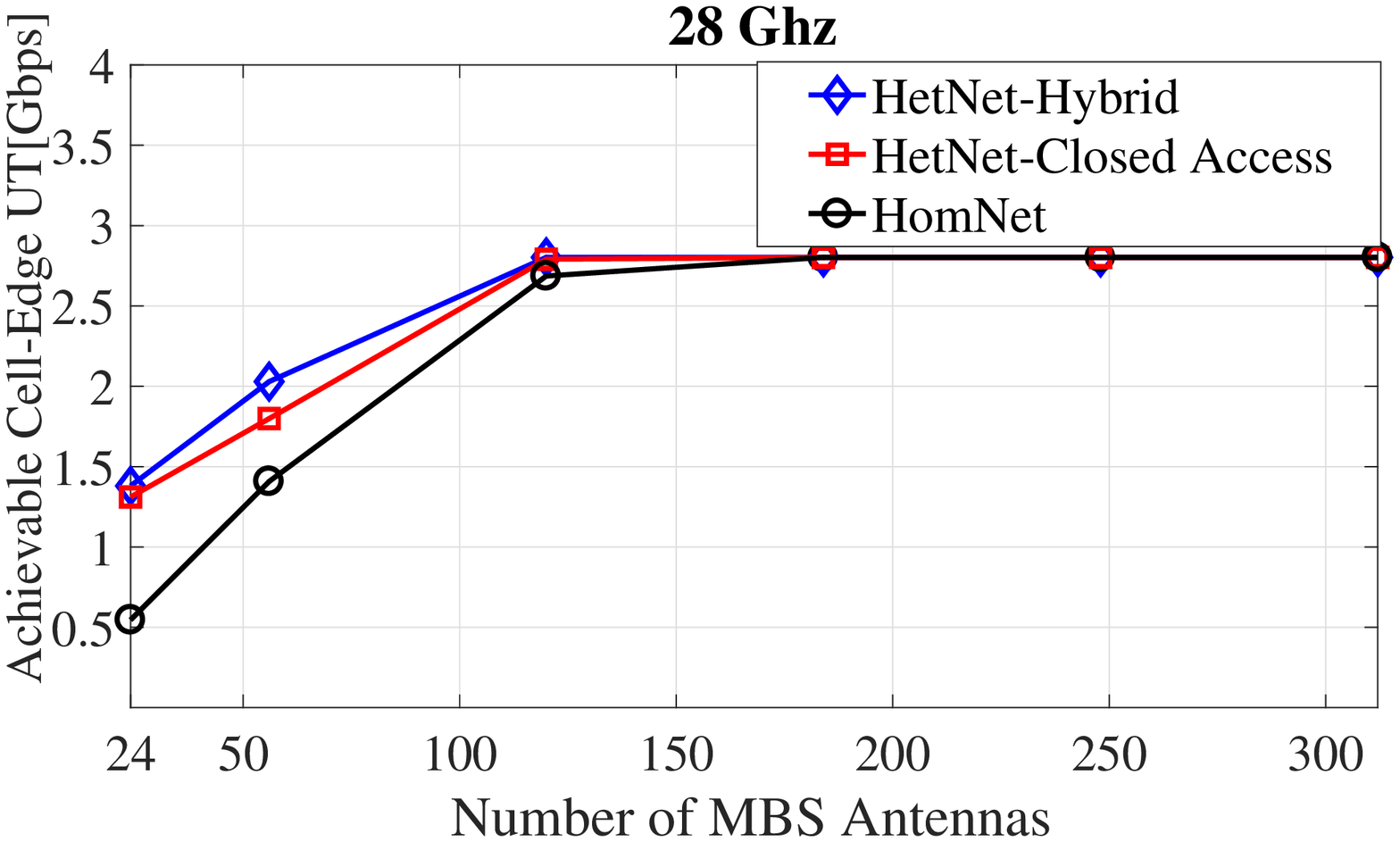} 
		\caption{Achievable cell-edge UT versus $\emph{N}$, when $\emph{K} = 12$.} \label{Cell-Edge-UEvsAntennas}
	\end{minipage}
\end{figure*}

For a given number of UEs and SCs, we show the backhaul impact by varying the number of MBS antennas (MIMO gain). We also increase the number of SCs antennas, $\emph{N}_{s}$, from 4 to 48. Here, we set the network area to $0.5$ by $0.5$ $\text{km}^{2}$, and consider $4$ SCs and $8$ MUEs. From the antenna theory~\cite{2005antenna}, the beamforming gain is logarithmically proportional to the number of antennas, and thus, as the number of antennas goes to infinity, the beamforming gain diminishes. The avgUT and the cell-edge UT as a function of the number of MBS antennas are shown in Fig.~\ref{AvgUEvsAntennas} and Fig.~\ref{Cell-Edge-UEvsAntennas}, respectively. For a not-so-large number of MBS antennas, our proposed algorithm $\textbf{HetNet-Hybrid}$ yields higher avgUT and cell-edge UT as compared to both baselines.  For
large number of antennas the MUE’s choice of associating with the near-by SCs or the MBS yields similar
payoffs, the gain of Massive MIMO by smart beamforming saturates. Hence, our proposed algorithm $\textbf{HetNet-Hybrid}$ and $\textbf{HetNet-Closed Access}$~\cite{Vu2016} are tending to be the same as the number of antennas grows large. In Fig.~\ref{Cell-Edge-UEvsAntennas}, the performance of cell-edge
UE throughput (cell-edge UT) of all schemes tends to be the same, when the number of antennas increases. Under the worst channel propagation of the cell-edge users, the performance of cell-edge users could not improve further, since all the network resources (transmit power and antennas) need to be shared among all
UEs in order to maximize the total network utility.

\subsection{~~~~Wireless Backhaul Impact versus Transmit Power Levels at different Frequency Bands}
\label{SubEvaluation3}
\begin{figure*}[!t]
\centering
\begin{minipage}{3in}
    \centering
    \includegraphics[width=0.95\textwidth, scale=0.5]{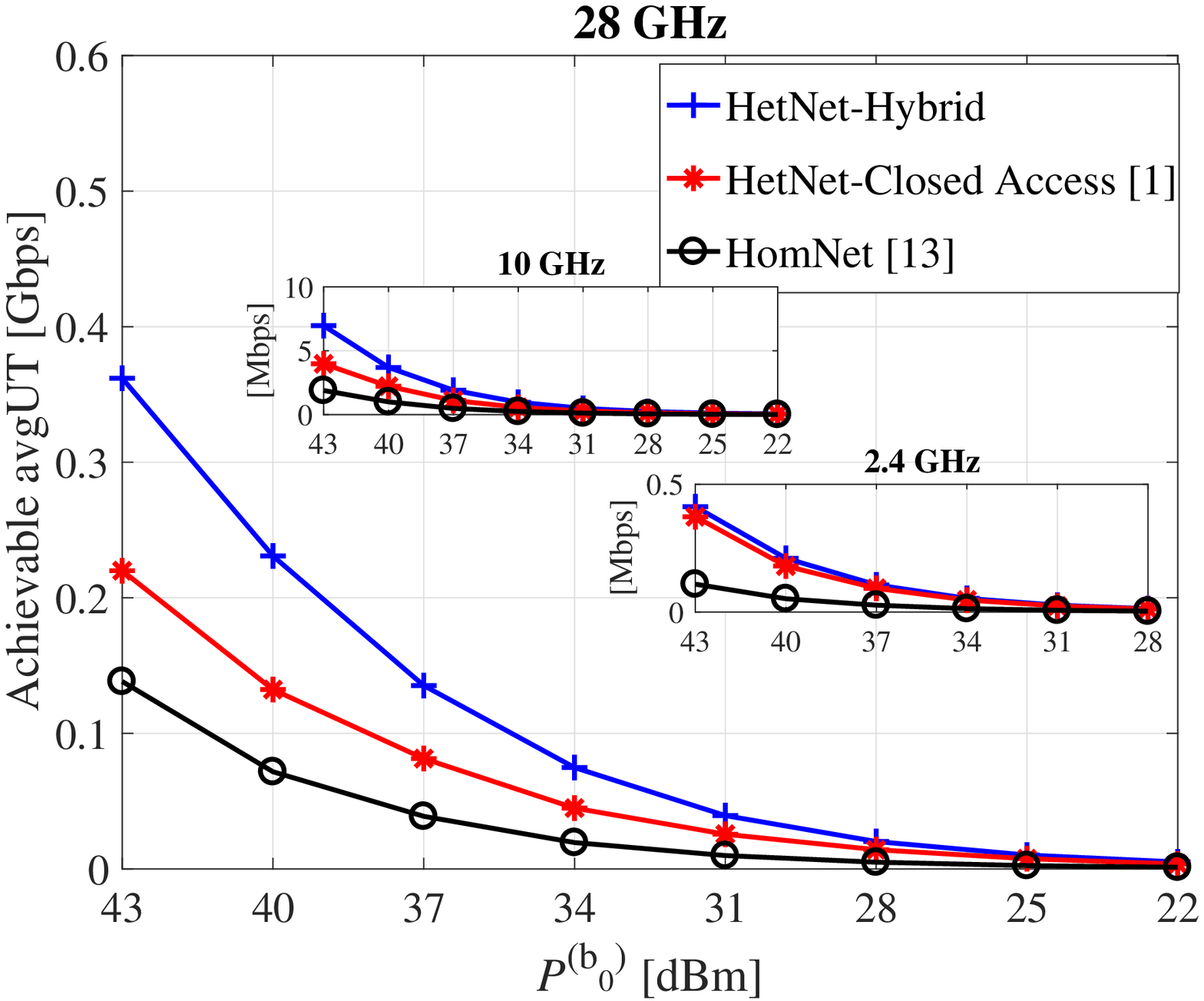}
    \caption{Achievable avgUT versus $P^{(b_0)}$ at $28$, $10$, and $2.4$ GHz, when $\emph{S} = 45$ per $\text{km}^{2}$, $\emph{K} = 3 \times \emph{S}$, $\emph{N} =  \emph{K}$. } \label{AvgUE}
 \end{minipage}
  \quad
 \begin{minipage}{3in}
    \centering
    \includegraphics[width=0.95\textwidth, scale=0.5]{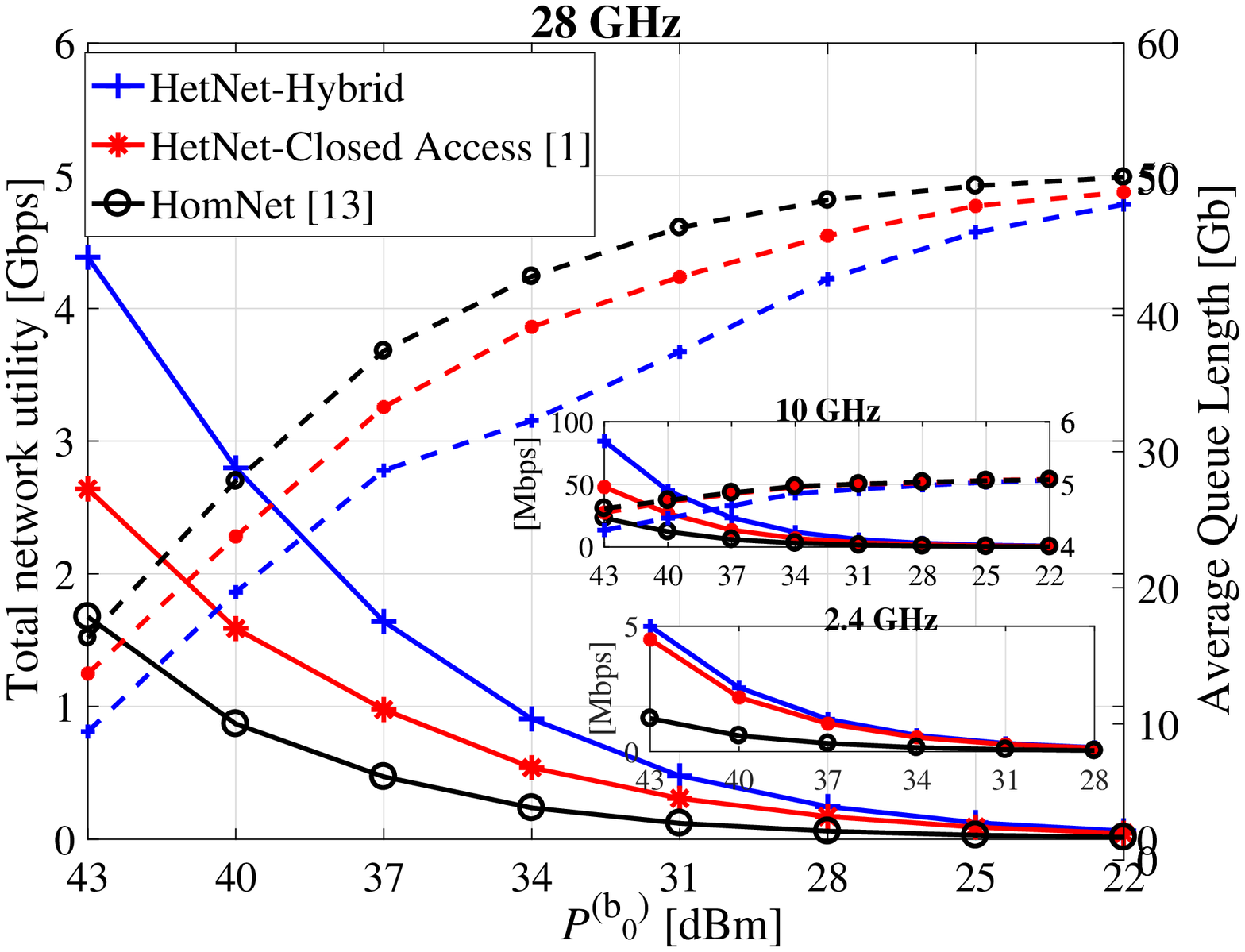} 
    \caption{TNU (``solid line") and network queue length (``dashed line") versus $P^{(b_0)}$ at $28$, $10$, and $2.4$ GHz, when $\emph{S} = 45$ per $\text{km}^{2}$, $\emph{K} = 3 \times \emph{S}$, $\emph{N} =  \emph{K}$.}\label{TotalUti}
 \end{minipage}
\end{figure*}

We also report the avgUT and the  total network utility (TNU) along with the average queue length (``dashed line") as a function of the MBS maximum transmit power at different frequency bands ($28$ GHz, $10$ GHz, and $2.4$ GHz) in Fig.~\ref{AvgUE} and Fig.~\ref{TotalUti}, respectively. In particular we consider the number of SCs is $\emph{S} = 45$ per $\text{km}^2$, and the number of MUEs $\emph{M}$ is twice the number of SCs $\emph{S}$. The number of MBS antennas is set to $\emph{N} = \emph{K}$, while the number of antennas at SCs $\emph{N}_{s} +1$ is set to $5$. Due to insufficient number of antennas at the MBS to simultaneously serve all MUEs and SCs and to alleviate the interference, offloading from the MBS to SCs helps to associate more UEs to the BSs. In this case the TNU is low, since the number of MBS antennas is reduced by half as compared to the impact of MBS antennas cases. As decreasing the maximum transmit power at the MBSs, $\textbf{HetNet-Hybrid}$ outperforms $\textbf{HetNet-Closed Access}$~\cite{Vu2016}, there is an inflexion point where the performance of $\textbf{HetNet-Hybrid}$ is close to that of $\textbf{HetNet-Closed Access}$~\cite{Vu2016} when the transmit power level is $25$ dBm, $31$ dBm, and $37$ dBm at $28$ GHz, $10$ GHz, and $2.4$ GHz, respectively. It can be observed that at higher frequency bands FD-enabled SCs work better at open access mode than closed access mode under the same transmit power budget. When the maximum MBS transmit power is too small, the performance of $\textbf{HetNet-Hybrid}$ and $\textbf{HetNet-Closed Access}$~\cite{Vu2016} is very closed to that of $\textbf{HomNet}$~\cite{wagner2012l}.

\begin{figure*}[!t]
\centering
 \begin{minipage}{3in}
    \centering
    \includegraphics[width=0.95\textwidth, scale=0.5]{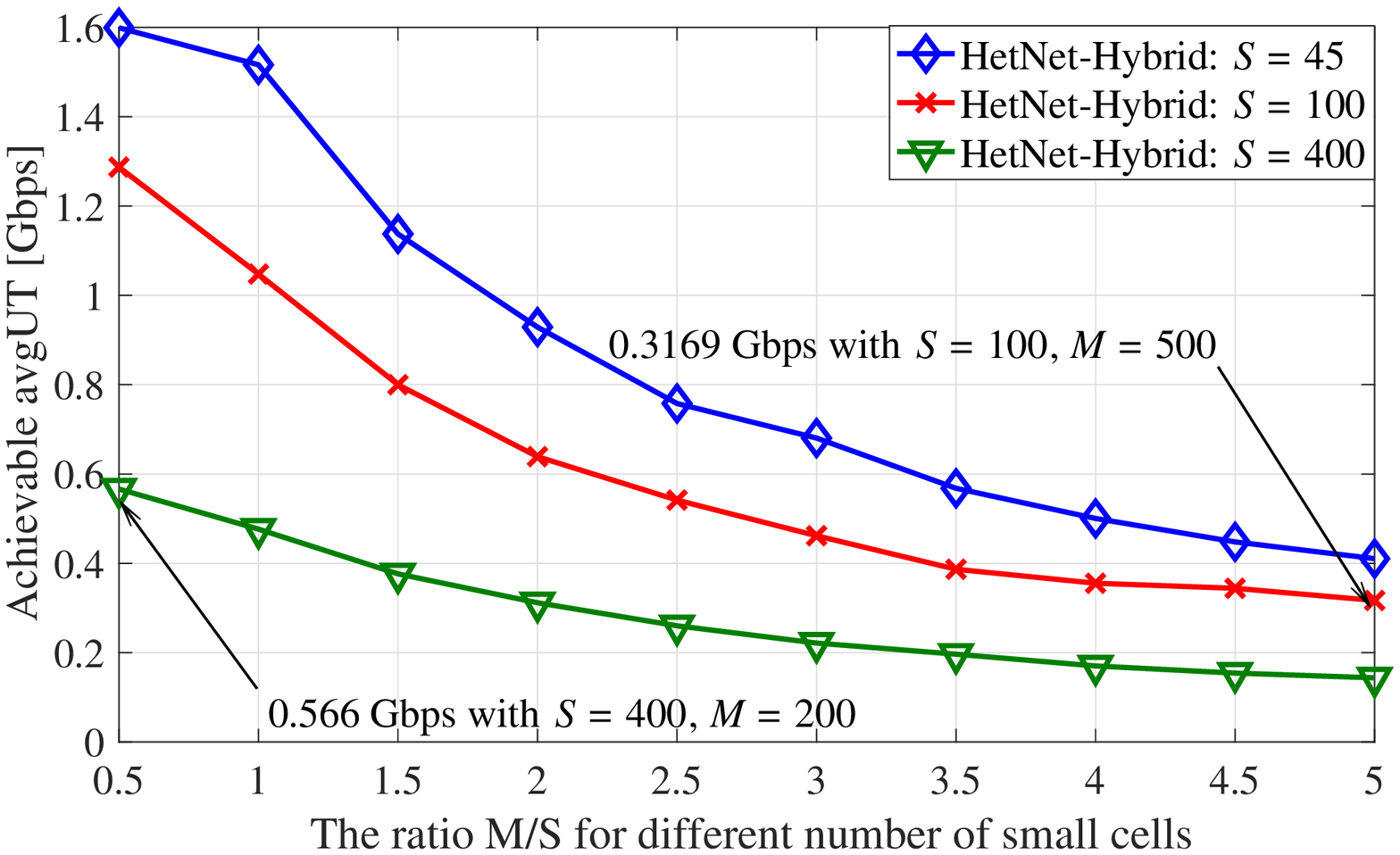}
    \caption{Achievable avgUT versus the ratio $\emph{M}$/$\emph{S}$ for different number of small cells.}
    \label{R3Q3}
 \end{minipage}
 \quad
 \begin{minipage}{3in}
    \centering
    \includegraphics[width=0.95\textwidth, scale=0.5]{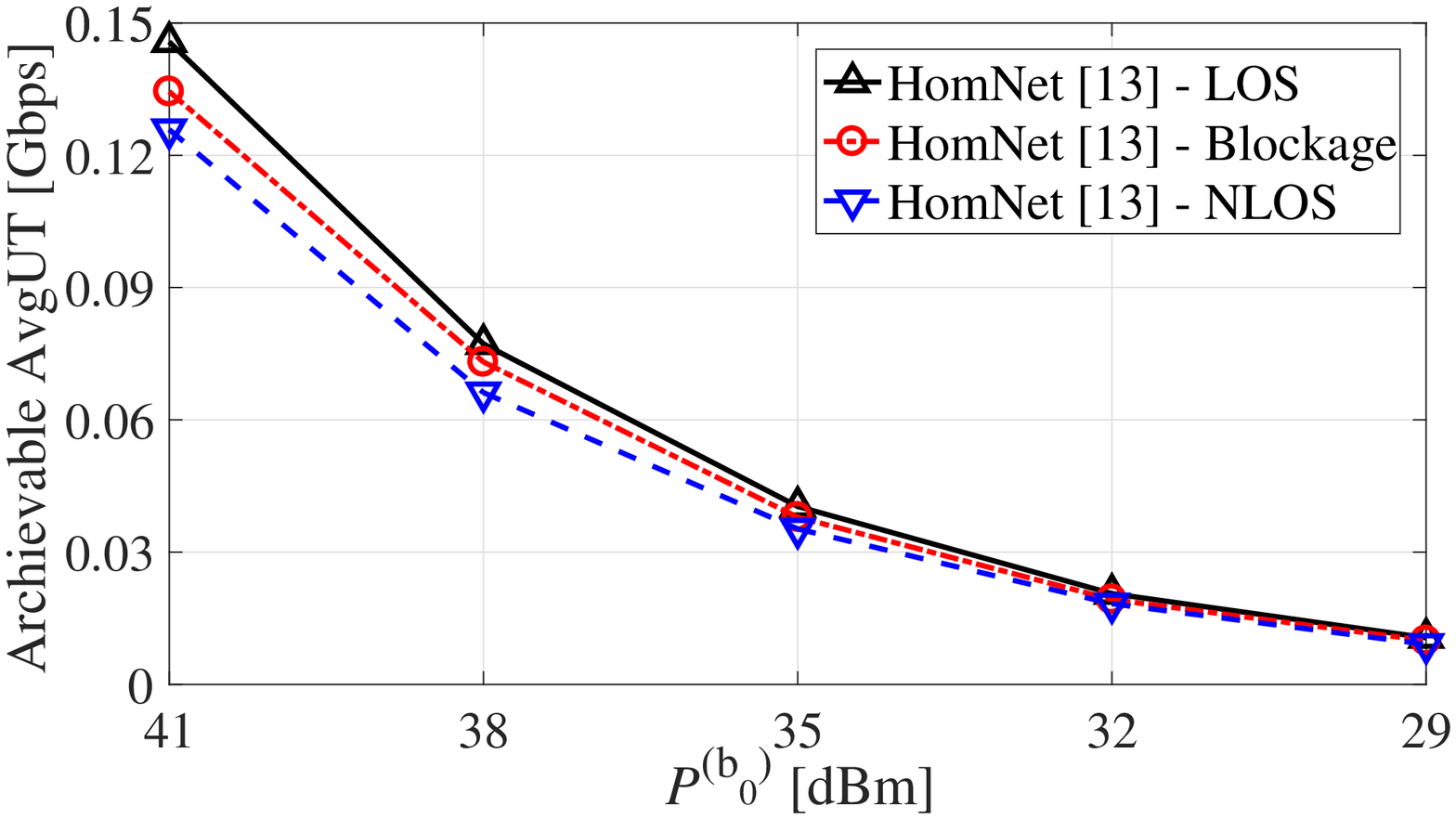}
    \caption{Effects of LOS, NLOS, and blockage channel models, when $\emph{K} = 3\times\emph{S}  = 12$ and  $\emph{N} = \emph{K}$.}
    \label{R3Q4}
 \end{minipage}
\end{figure*}

Moreover, in Fig.~\ref{R3Q3} we report the avgUT versus the ratio of number of MUEs to number of SCs, $\emph{M}/\emph{S}$, under different sets of SCs. Here, the number of SCs per $\text{km}^{2}$ is set to $45$, $100$, and $400$ representing the network density from sparse to dense, while the ratio $\emph{M}/\emph{S}$ is varying from $0.5$ to $5$. It can be observed that under the same total number of UEs, i.e., $\emph{K}=600$, deploying denser number of SCs with $\emph{S}=400$ and $\emph{M}=200$ obtains avgUT of $0.566$ Gbps, which is higher than $0.3169$ Gbps for a system with less number of SCs $\emph{S}=100$ and $\emph{M}=500$.

We have used the LOS channel model to make the performance evaluation such that  the probability of obtaining LOS is very high. We now report the impact of channel models on massive MIMO system operating at $28$ GHz mmWave frequency band. Beside the LOS and non-LOS (NLOS) channel states, there exists another channel state called blockage state, which is modeled as a distance-dependence probability state where the channel is either LOS or NLOS by using the stochastic model~\cite{Bai2014}. In Fig.~\ref{R3Q4}, the performance gap between LOS and blockage channel models is shown versus the maximum transmit power.

\subsection{~~~~Convergence} %
\begin{figure*}[!t]
\centering
 \begin{minipage}{3in}
    \centering
    \includegraphics[ scale=0.36]{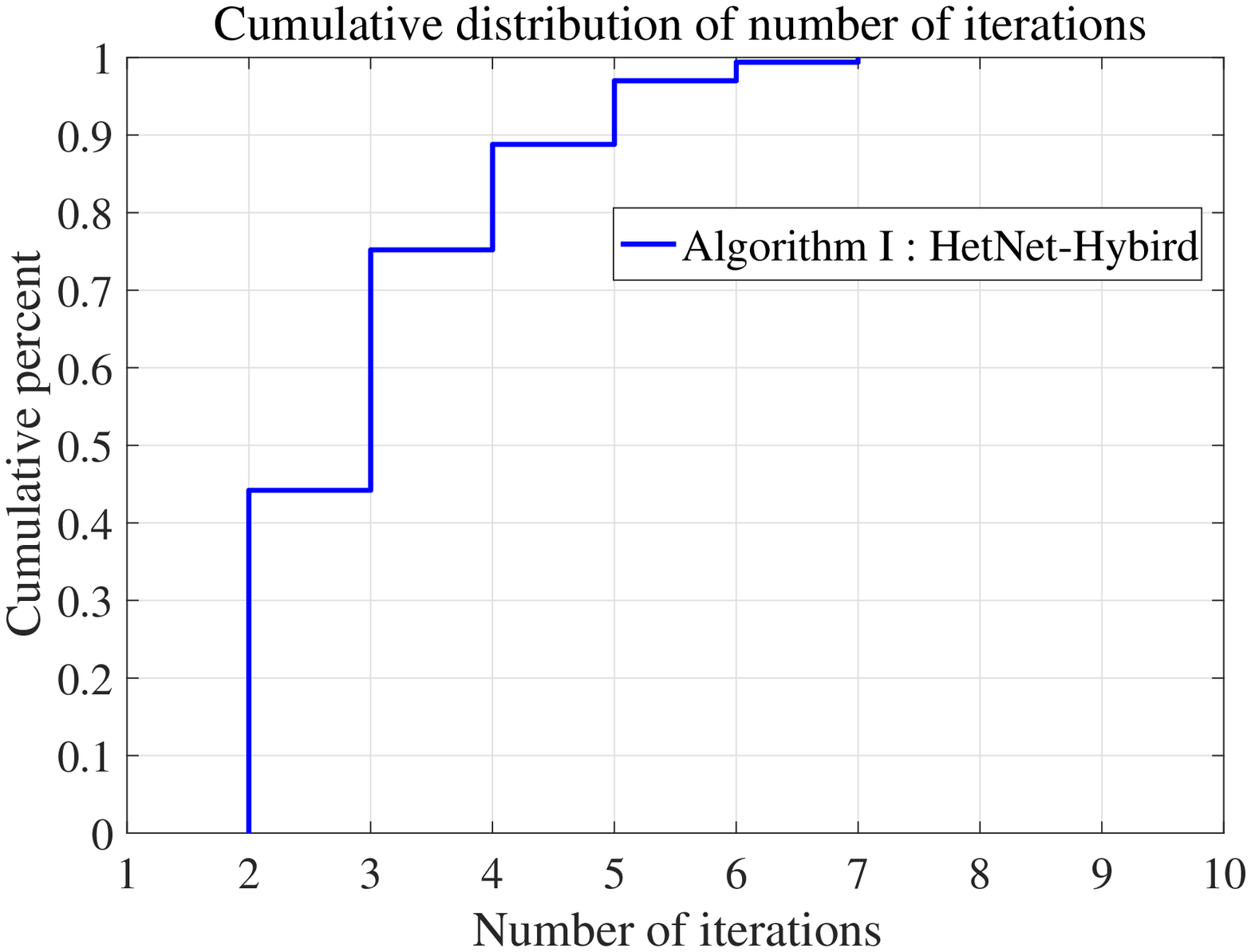}
    \caption{The CDF of convergence of $\textbf{Algorithm~\ref{algLB}}$.}
    \label{Convergence}
 \end{minipage}
\quad
  \begin{minipage}{3in}
     \centering
    \includegraphics[scale=0.36]{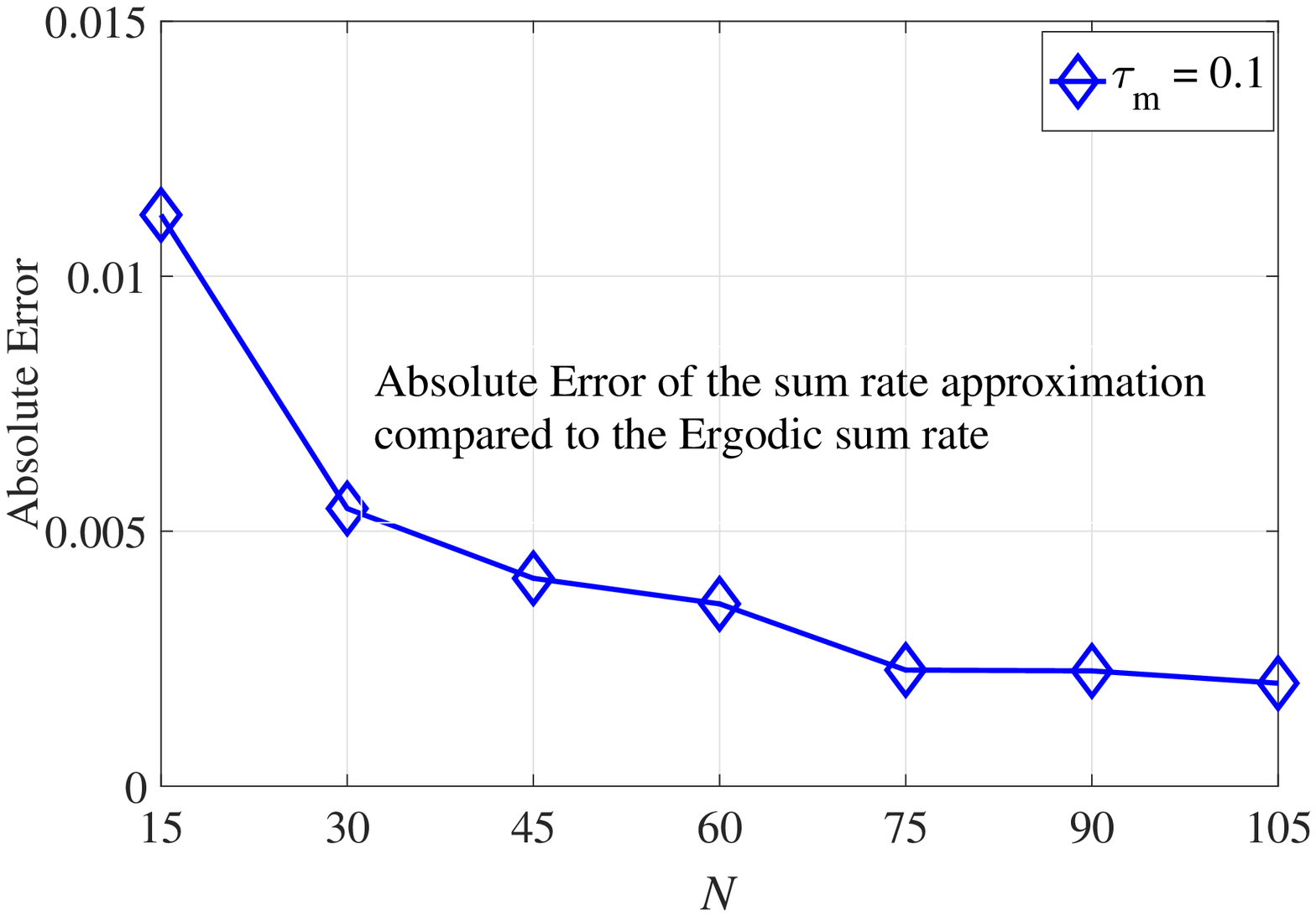}
    \caption{Validation of the approximation of the closed-form expression, when $\emph{K} = 12$ and  $\emph{K} = 3 \times \emph{S}$.}
    \label{Validated}
 \end{minipage}
\end{figure*}

\label{SubEvaluation3}
In Fig.~\ref{Convergence} we show the convergence behaviour of our approximated algorithm based on the SCA method when deploying our $\textbf{HetNet-Hybrid}$ algorithm. While the convergence analysis is provided in Appendix~\ref{ConvergenceAna}. Unlike other works, we plot the cumulative distribution of the number of iterations at which the $\textbf{Algorithm~\ref{algLB}}$ converges for all $t$. We observe that the probability that the number of iterations takes on a value less than or equal to 4 is $90\%$, which implies that our proposed algorithm only needs few iterations to converge.

We then validate the accuracy of the closed-form expression for the data rate by comparing the Ergodic sum rate $\emph{R}$, which is obtained by using the $\mathrm{SINR}$ from~(\ref{SINR-MUE-2}) and~(\ref{SINR-SC-2}) from simulations of i.i.d. Rayleigh block-fading channels, to the approximated sum rate $\tilde{\emph{R}}$, which obtained by using $\mathrm{SINR}$ from~(\ref{SINR-MUE-3}) and~(\ref{SINR-SC-3}). The sum rate is defined as the total sum of all user data rates. We define the absolute error as $\frac{\tilde{\emph{R}}-{\emph{R}}}{\tilde{\emph{R}}}$, then we plot the absolute error versus the number of MBS antennas, while the number of users is fixed to $\emph{K} = 12$. As can be seen in Fig.~\ref{Validated}, the absolute error decreases as increasing number of MBS antennas. It means that closed-form expressions is more accurate when number of MBS antennas is higher than number of users, i.e., $\emph{N} \gg \emph{K}$.
The impact of the Lyapunov parameter $\nu$ on the achievable average network utility and queue backlog has been showed in our previous work~\cite{Vu2016}. It has been observed that the network utility is increasing
with $\mathcal{O}(1/\nu)$, while the network backlog linearly increases with $\mathcal{O}(\nu)$. Hence, choosing the value of $\nu$ will result in an $[\mathcal{O}(1/\nu), \mathcal{O}(\nu)]$ utility-queue backlog tradeoff, which leads to an utility-delay tradeoff~\cite{vu2017ultra}.
\subsection{Effect of Pilot Training and Channel Aging}
\label{SubEvaluation5}
In this subsection we study the effect of pilot training and channel aging in $5$G Massive MIMO system as we consider a large number of antennas and users. We assume that each coherence interval consists of three phases:
uplink (UL) training, DL payload data transmission, and UL payload data transmission. During the UL training phase, the users send pilot sequences to the BSs and each BS estimates the channels. The estimate channels are
used to precode the transmit signals in the DL. Let $\emph{T}_\mathrm{ci}$ and  $\tau_{\mathrm{td}}$ denote the length of the coherence interval and the UL training duration, where the subscripts $\mathrm{ci}$ and
$\mathrm{td}$ stand for ``coherence interval" and  ``training duration", respectively. Basically, we have the length of pilot sequences is less than the length of the coherence interval, i.e., $\tau_{\mathrm{td}}  <
\emph{T}_\mathrm{ci}$. If the length of pilot sequences $\tau_{\mathrm{td}}$ is greater or equal to the number of user $\emph{K}$, $\tau_{\mathrm{td}} \geq \emph{K}$, to achieve the estimate channel the pilot assignment is
chosen pairwisely orthogonal. However, if $\tau_{\mathrm{td}} < \emph{K}$, then channel estimate is degraded due to non-orthogonal pilot signals that leads to the pilot contamination effect. In this work we consider the imperfect CSI, which is due to channel estimation errors during
the UL training and the coherence interval $\emph{T}_\mathrm{ci}$, while the channel reciprocity is perfect for UL and DL. In addition, the channel aging is also a very important issue needed to be addressed in Massive MIMO
systems, and the Massive MIMO systems are most suitable for static users with not-too-fast movement. For simplicity, we consider the problem of channel aging by the impact of channel estimate error factor $\tau_k$ as shown
in~(\ref{Channel-Er}). We next show the relation between the channel estimate error and the length of pilot sequences. Assume that each user will transmit the orthogonal pilot signal to the MBS during the UL training in
which
$\tau_{\mathrm{td}} \geq \emph{K}$ and the MBS receives the pilot signals simultaneously. Follow the analysis in~\cite{wagner2012l}, the channel estimate error is $\tau_{k}^{2}  = \frac{1} {1 + \tau_{\mathrm{td}} \rho^\mathrm{ul}}$, where $\rho^\mathrm{ul}$ is defined as the UL signal-to-noise ratio of user $k$, $\rho^\mathrm{ul} = p^{\mathrm{ul}}_{k}/ \sigma_k^2$, here $p^{\mathrm{ul}}_{k}$ is the UL transmit power of user $k$.

Since our work considers only the performance of the DL, we then simplistically assume that the transmission time for DL and UL during the coherence time is identical$\footnote{Due to different traffic model for DL and UL,
the
ratio configuration for DL and UL transmission time will be varied.}$. Therefore, the closed-form expression of UT taking into account the impact of pilot training, coherence interval, and channel aging is
\begin{equation}
    r_{k}^{\mathrm{ci}}(\mathbf{\Lambda}|\mathbf{\Theta}) \xrightarrow{a.s.} \mathrm{B}  \frac{(1-\tau_{\mathrm{td}}/\emph{T}_\mathrm{ci})}{2}  \log \big( 1 +  \frac{\us_k^{(b_0)} p_{k}^{(b_0) } (1 - \frac{1} { 1 +
    \tau_{\mathrm{td}} \rho^\mathrm{ul}})}{1 + \textstyle \sum _{s = 1}^{\emph{S}} \op^{(b_s)} \EP_{k}^{(b_s)}} \big), \notag
\end{equation}
where $\mathrm{B}$ is the channel bandwidth. To set up the simulation parameters for this subsection all the transmit powers are normalized by dividing by the thermal noise power $\sigma_k$. At $28$ GHz we set the coherence
interval to $\emph{T}_\mathrm{ci} = 350$ with the coherence bandwidth of $10$ MHz and the coherence time of \SI{35}{\micro\second}.

\begin{figure}[!t]
     \centering
    \includegraphics[scale=0.4]{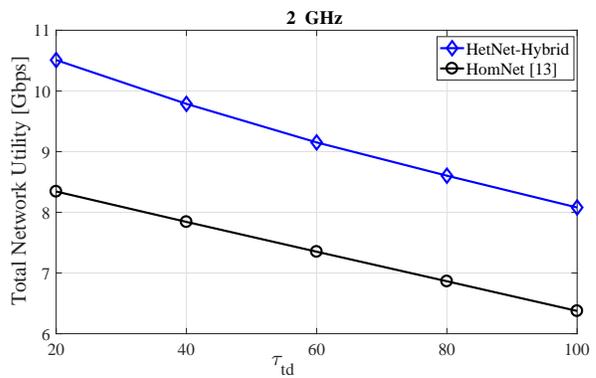} 
    \caption{Impact of pilot sequence length, when $\emph{K} = 12$, $\emph{N} =  2 \times \emph{K}$, and  $\emph{K} = 3 \times \emph{S}$.}\label{PilotImpact}
\end{figure}

We show the impact of pilot training on the total network utility of our proposed HetNet-Hybrid as compared to HomNet at $28$ GHz. To do that, we vary the length of pilot sequences from $20$ to $100$, while the number of pilot
training sequence is greater than number of users. As can be seen in Fig.~\ref{PilotImpact}, with increasing the pilot training duration, the total network utility for the DL is gradually degraded.
\section{~~~~~~~Conclusion}
\label{Conclusion}
We have studied the NUM by considering the problem of joint load balancing (user association and user scheduling) and interference management (beamforming design and power allocation) taking into account the dynamic wireless backhaul, traffic demand, and imperfect CSI in $5$G HetNets. The problem of load balancing has taken into account the wireless backhaul capacity in order to reduce the load of the MBS while avoiding the bottleneck problem at the FD-enabled SCs. By utilizing a very large number of antennas at MBS we design a hierarchical precoder in order to mitigate both cross-tier and co-tier interference in HetNets. We aim to maximize a network utility function of the total time-average data rates subject to the backhaul dynamic and network stability in the presence of imperfect CSI. We have exploited from RMT to obtain a closed-form expression of the original problem in a large system regime. By applying the stochastic optimization, the problem is then decoupled decoupled into dynamic scheduling of MUEs, backhaul provisioning of in-band FD-enabled SCs, and offloading UEs to in-band FD-enabled SCs as a function of interference, number of antennas, and backhaul loads. We have provided the performance analysis and validation of our proposed algorithm, which states that there exists an $[\mathcal{O}(1/\nu), \mathcal{O}(\nu)]$ utility-queue backlog tradeoff. Via numerical results, we have show that our proposed algorithm outperforms the baselines with respect to the number of SCs per $\text{km}^{2}$, the number of MBS antennas, and the MBS transmit power levels at different frequency bands. Interestingly, we find that even at lower frequency band the performance of open access small cell is close to that of closed access at some operating points, the open access full-duplex small cell still yields higher gain as compared to the closed access at higher frequency bands. With increasing the small cell density or the wireless backhaul quality, the open access full-duplex small cells outperform and achieve $5.6\times$ gain in terms of cell-edge performance as compared to the closed access ones in ultra-dense networks with $350$ small cell base stations per $\text{km}^2$.
\appendices
\section{~~Convergence Analysis for $\textbf{Algorithm~\ref{algLB}}$ }
\label{ConvergenceAna}
Next, we establish a convergence result for $\textbf{Algorithm~\ref{algLB}}$ based on the SCA method, since the original problem~(\ref{Original-US}) has a non-convex objective function~(\ref{eq-OUS1}) subject to
non-convex constraint~(\ref{eq-OUS7}). By using the SCA method, we replace the original nonconvex problem~(\ref{Original-US}) by a strongly convex problem~(\ref{Optimal-US}). We will briefly describe the convergence here for the sake of completeness since it was
studied in~\cite{beck2010seq, tran2012}. We assume that the $\textbf{Algorithm~\ref{algLB}}$ obtains the solution of problem~(\ref{Optimal-US}) at iteration ${i +1}~{\mathrm{th}}$. The updating rule in $\textbf{Algorithm~\ref{algLB}}$
ensures that the optimal values $\mathbf{\Lambda}^{o(i)}$, $\lambda_{ks}^{(i)}$, and $\hat{\iota}^{(i)}_{s}$ at iteration $i$ satisfy all constraints in~(\ref{Optimal-US}) and are feasible to the optimization problem at
iteration $i+1$. Hence, the objective obtained in the $i+1\mathrm{st}$ iteration is less than or equal to that in the in the $i\mathrm{th}$ iteration, since we minimize the convex function. In other words,
$\textbf{Algorithm~\ref{algLB}}$ yields a non-increasing sequence. Due to antenna and interference constraints, the objective is bounded, and thus $\textbf{Algorithm~\ref{algLB}}$ converges to some local optimal solution
of~(\ref{Optimal-US}). Moreover, $\textbf{Algorithm~\ref{algLB}}$ produces a sequence of points that are feasible for the original problem~(\ref{Original-US}) and this solution is satisfied the KKT condition of the original
problem~(\ref{Original-US}) as discussed in~\cite{beck2010seq, tran2012}.
\section{~Performance Analysis}
\label{Analys}
$\textbf{Theorem~\ref{Theo1}}$ is provided to show the performance analysis of network utility maximization based on Lyapunov framework and prove that the queues are stable.
\begin{theorem}\label{Theo1}[Optimality] Assume that all queues are initially empty. For arbitrary arrival rates, the operation mode and load balancing is chosen to satisfy~(\ref{FinalEQ}) and the rate regime. For a given
constant $\chi \geq 0$, the network utility maximization with any $\nu > 0$ provides the following utility performance with $\chi-\textit{approximation}$
\[
{f}_0 \geq {f}^{\ast}_{0} - \frac{ \Psi + \chi}{\nu},
\]
where ${f}^{\star}_{0}$ is the optimal network utility over the rate regime.
\end{theorem}
$\textit{Proof:}$ To prove the $\textbf{Theorem~\ref{Theo1}}$, we first prove the queues are bounded. Let $\pi_k$ denote the largest right derivative of $f( \bar{r}_{k} )$, the Lyapunov framework can guarantee the following
strong stability of the virtual queues and the network queues.
\begin{equation}\label{Bound1}
{Q}_{k}(t) \leq \nu \omega_k(t) \pi_k + 2 a_{k}^{\mathrm{max}},
\end{equation}
\begin{equation}\label{Bound2}
{Y}_{k}(t)  \leq  \nu \omega_k(t) \pi_k + a_{k}^{\mathrm{max}},
\end{equation}
\begin{equation}\label{Bound3}
{D}_{s}(t)  \leq  \nu \omega_s(t) \pi_s + a_{s+M}^{\mathrm{max}}.
\end{equation}
Here we first prove the bound of the virtual queues, and then the bound of the network queues are proved similarly. Suppose that all queues are initially empty, this clearly holds for $t = 0$. Suppose these inequalities hold
for some $t > 0$, we need to show that it also holds for $t + 1$.

From~(\ref{queueD}) and~(\ref{queueY}), if ${Y}_{k}(t)  \leq  \nu \omega_k(t) \pi_k$ and ${D}_{s}(t)  \leq  \nu \omega_s(t) \pi_s$ then ${Y}_{k}(t+1)  \leq  \nu \omega_k(t) \pi_k + a_{k}^{\mathrm{max}}$ and ${D}_{s}(t+1) \leq  \nu \omega_s(t) \pi_s + a_{s+M}^{\mathrm{max}}$ and the bound holds for $t+1$ due to the arrival rate constraint ${r}_{k}(t)  \leq a_{k}^{\mathrm{max}}$ and ${r}_{s}(t)  \leq a_{s}^{\mathrm{max}}$. Else, if  ${Y}_{k}(t)
\geq  \nu \omega_k(t) \pi_k$ and ${D}_{s}(t)  \geq  \nu \omega_s(t) \pi_s$; since the value of auxiliary variables is determined by maximized $\textstyle \sum _{k = 1}^{\emph{K}} {Y}_{k}(t) \varphi_{k}(t) + \textstyle \sum
_{s = 1}^{\emph{S}} {D}_{s}(t) {\varphi}_{s+M}(t) -  \nu    {f}_{0}(\boldsymbol{\varphi}(t))$, $\boldsymbol{\varphi}(t)$ is then forced to be zero. From~(\ref{queueY}) and~(\ref{queueD}), ${Y}_{k}(t+1)$ and ${D}_{s}(t+1)$
are bounded by ${Y}_{k}(t)$ and ${D}_{s}(t)$, respectively. Since the virtual queues are bounded for $t$, we have the following inequalities
\begin{equation}\label{Bound22}
{Y}_{k}(t+1)  \leq {Y}_{k}(t)  \leq  \nu \omega_k(t) \pi_k + a_{k}^{\mathrm{max}},
\end{equation}
\begin{equation}\label{Bound32}
{D}_{s}(t+1)  \leq {D}_{s}(t)  \leq  \nu \omega_s(t) \pi_s + a_{s+M}^{\mathrm{max}}.
\end{equation}
Hence, the bounds of the virtual queues hold for all $t$. Similarly, we show that the network queue~(\ref{Bound1}) holds for all $t$. It clearly holds for $t = 0$. We assume that~(\ref{Bound1}) holds for $t >0$, we now prove
it holds for $t+1$. Note that from~(\ref{queueQ}) and ~(\ref{queueY}) we have ${Q}_{k}(t+1) \leq {H}_{k}(t+1) + a_k(t)$. Moreover, we just proved that ${H}_{k}(t+1) \leq \nu \omega_k(t) \pi_k + a_{k}^{\mathrm{max}}$ then we
have ${Q}_{k}(t+1) \leq \nu \omega_k(t) \pi_k + 2 a_{k}^{\mathrm{max}}$ and the network bound holds for $t+1$.

We have established the network bounds, we are going to show the utility bound. Since our solution of (\ref{eq:Obj-Formulate-1}) is to minimize the Lyapunov drift and the objective function every time slot $t$, we have
the following inequality
\begin{equation}\label{Proof1}
\begin{aligned}
&\mathbf{\Delta}(\mathbf{\Sigma}(t)) -\nu \mathbb{E}[{f}_0 ( \boldsymbol {\varphi}(t) )] \leq ~
\\&~~~~~~\Psi -\nu \mathbb{E}[{f}_0 ( \boldsymbol {\varphi}^{\ast}(t) )] +  \textstyle \sum _{k = 1}^{\emph{K}} {Q}_{k}(t) \mathbb{E} \Big[  {a}_{k}(t) - {r}_{k}^{\ast}(t) | \mathbf{\Sigma}(t) \Big ]
\\&~~~~~~+ \textstyle \sum _{k = 1}^{\emph{K}} {Y}_{k}(t) \mathbb{E} \Big[ \varphi_{k}^{\ast}(t) - r_{k}^{\ast}(t) | \mathbf{\Sigma}(t) \Big]
\\&~~~~~~+ \textstyle \sum _{s = 1}^{\emph{S}} {D}_{s}(t)\mathbb{E} \Big[  {\varphi}^{\ast}_{s+M}(t)  - \op^{(b_s) \ast}(t) {r}_{s}^{\sue_{s}{\ast}}(t) |\mathbf{\Sigma}(t) \Big ], \nonumber
\end{aligned}
\end{equation}
where $\boldsymbol {\varphi}^{\ast}(t), \op^{(b_s) \ast}(t)$, and $r_{k}^{\ast}(t)$ are the optimal values of the problem~(\ref{FinalEQ}). Since the queues are bounded, for given $\chi \geq 0$, obtaining
\begin{equation}\label{Proof2}
\begin{split}
\mathbf{\Delta}(\mathbf{\Sigma}(t)) -\nu \mathbb{E}[{f}_0 ( \boldsymbol {\varphi}(t) )] \leq \Psi -\nu \mathbb{E}[{f}_0 ( \boldsymbol {\varphi}^{\ast}(t) )] +  \chi. \nonumber
\end{split}
\end{equation}
By taking expectations of both sides of the above inequality and choosing $\mathbf {r}^{\ast}(t) = \boldsymbol {\varphi}^{\ast}(t)$, it yields for all $t \geq 0$,
\begin{equation}\label{Proof3}
\begin{split}
\mathbb{E} \big [ {L}(\mathbf{\Sigma}(t+1))  - &{L}(\mathbf{\Sigma}(t))| \mathbf{\Sigma}(t) \big] - \nu \mathbb{E}[{f}_0 ( \boldsymbol {\varphi}(t) )] \leq \\ &~\Psi +  \chi  -\nu \mathbb{E}[{f}_0 ( \mathbf
{r}^{\ast}(t) )].\nonumber
\end{split}
\end{equation}
By taking the sum over $\tau = 0, \ldots, t-1$ and dividing by $t$, (using the fact that ${f}_0 ( \mathbf {r}^{\ast}(t)) = {f}_{0}^{\ast}$), yielding
\begin{equation}\label{Proof4}
\begin{split}
\frac{\mathbb{E} \big [ {L}(\mathbf{\Sigma}(t+1))  - {L}(\mathbf{\Sigma}(0))| \mathbf{\Sigma}(t) \big]}{t} & - \frac{\nu}{t} \sum_{\tau=0}^{t-1} \mathbb{E}[{f}_0 ( \boldsymbol {\varphi}(t) )] \leq \\&  ~\Psi +  \chi -\nu
{f}_{0}^{\ast}.
\end{split}
\end{equation}
By using the fact that ${L}(\mathbf{\Sigma}(t+1)) \geq 0$ and ${L}(\mathbf{\Sigma}(0)) = 0$, and applying Jensen's inequality in the concave function and rearranging term yields
\[
{f}_{0}( \boldsymbol {\varphi}(t) ) \geq~{f}_{0}^{\ast} - \frac{ \Psi + \chi}{\nu}.
\]
Since the network utility function is a non-decreasing concave function, the auxiliary variable is chosen to satisfy $r_k(t) \geq \varphi_k(t)$. Hence ${f}_{0}( \mathbf{r}(t) ) \geq {f}_{0}( \boldsymbol {\varphi}(t) )
\geq~{f}_{0}^{\ast} - \frac{ \Psi + \chi}{\nu}$, which means that the solution is closed to the optimal as increasing $\nu$. Which completes the proof of the $\textbf{Theorem~\ref{Theo1}}$. Hence, there exists an
$[\mathcal{O}(\nu), \mathcal{O}(1/\nu)]$ utility-queue length tradeoff, which leads to an utility-delay balancing.

We now prove that all queues are stable by using the $\textbf{Definition~\ref{QueueDef}}$, the bound (\ref{Proof4}) can be rewritten as
\[
\mathbf{\Delta}(\mathbf{\Sigma}(t)) \leq~C,
\]
where $C$ is any constant that satisfies for all $t$ and $\mathbf{\Sigma}(t)$: $C \geq  \Psi +  \chi  -\nu ({f}_{0}^{\ast} - \mathbb{E}[{f}_{0}( \boldsymbol {\varphi}(t) )] )$. By using the definition of the Lyapunov
drift and taking an expectation, obtaining
\[
\mathbb{E} \big [ {L}(\mathbf{\Sigma}(t)) \big] \leq~Ct.
\]
As the definition of the Lyapunov function ${L}(\mathbf{\Sigma}(t))$ we have
\[
\mathbb{E}[ {Q}_k(t) ]^{2}, \mathbb{E}[ {H}_k(t) ]^{2}, \mathbb{E}[ {D}_s(t) ]^{2} \leq~2Ct.
\]
Dividing both sides by $t^2$, and taking the square roots shows for all $t>0$:
\[
\frac{\mathbb{E}[ {Q}_k(t) ]}{t}, \frac{\mathbb{E}[ {H}_k(t) ]}{t}, \frac{\mathbb{E}[ {D}_k(t) ]}{t} \leq~\sqrt{ \frac{2C}{t}}.
\]
As $t \rightarrow \infty$, taking the limit, we prove the queues are stable.

\section*{Acknowledgment}
The authors would like to acknowledge colleagues: Kien Giang Nguyen, Mohammed ElBamby, and Petri Luoto for helpful discussions on the paper.

\bibliographystyle{IEEEtran}
\bibliography{my_reference_journal1}

\begin{thebibliography}{10}
\providecommand{\url}[1]{#1}
\csname url@samestyle\endcsname
\providecommand{\newblock}{\relax}
\providecommand{\bibinfo}[2]{#2}
\providecommand{\BIBentrySTDinterwordspacing}{\spaceskip=0pt\relax}
\providecommand{\BIBentryALTinterwordstretchfactor}{4}
\providecommand{\BIBentryALTinterwordspacing}{\spaceskip=\fontdimen2\font plus
\BIBentryALTinterwordstretchfactor\fontdimen3\font minus
  \fontdimen4\font\relax}
\providecommand{\BIBforeignlanguage}[2]{{%
\expandafter\ifx\csname l@#1\endcsname\relax
\typeout{** WARNING: IEEEtran.bst: No hyphenation pattern has been}%
\typeout{** loaded for the language `#1'. Using the pattern for}%
\typeout{** the default language instead.}%
\else
\language=\csname l@#1\endcsname
\fi
#2}}
\providecommand{\BIBdecl}{\relax}
\BIBdecl

\bibitem{Vu2016}
T.~K. Vu, M.~Bennis, S.~Samarakoon, M.~Debbah, and M.~Latva-aho, ``{Joint
  In-Band Backhauling and Interference Mitigation in 5G Heterogeneous
  Networks},'' in \emph{Proceedings of 22th European Wireless Conference},
  Oulu, Finland, May 2016, pp. 1--6.

\bibitem{Nokia2011}
{Nokia~Siemens~Networks}, ``{2020: Beyond 4G Radio Evolution for the Gigabit
  Experience},'' White Paper, Nokia Siemens Networks, 2011.

\bibitem{marzt2010non}
T.~L. Marzetta, ``Noncooperative cellular wireless with unlimited numbers of
  base station antennas,'' \emph{IEEE Transactions on Wireless Communications},
  vol.~9, no.~11, pp. 3590--3600, 2010.

\bibitem{vu2015cooperative}
T.~K. Vu, S.~Kwon, and S.~Oh, ``{Cooperative Interference Mitigation Algorithm
  in Heterogeneous Networks},'' \emph{IEICE Transactions on Communications},
  vol.~98, no.~11, pp. 2238--2247, 2015.

\bibitem{li2015small}
L.~Boyu~{et al.}, ``{Small cell in-band wireless backhaul in massive
  Multiple-Input Multiple-Output systems},'' in \emph{IEEE International
  Conference on Communications}, London, UK, June 2015, pp. 1838--1844.

\bibitem{hur2013millimeter}
S.~Hur~{et al.}, ``Millimeter wave beamforming for wireless backhaul and access
  in small cell networks,'' \emph{IEEE Transactions on Communications},
  vol.~61, no.~10, pp. 4391--4403, 2013.

\bibitem{S2014if}
L.~Sanguinetti, A.~Moustakas, and M.~Debbah, ``{Interference management in 5G
  reverse TDD HetNets: A large system analysis},'' \emph{IEEE Journal on
  Selected Areas in Communications}, vol.~33, pp. 1187--1200, 2015.

\bibitem{2013user}
Q.~Ye~{et al.}, ``User association for load balancing in heterogeneous cellular
  networks,'' \emph{IEEE Transactions on Wireless Communications}, vol.~12,
  no.~6, pp. 2706--2716, 2013.

\bibitem{2015user}
D.~Bethanabhotla~{et al.}, ``{Optimal User-Cell Association for Massive MIMO
  Wireless Networks},'' \emph{IEEE Transactions on Wireless Communications},
  vol.~15, no.~3, pp. 1835--1850, March 2016.

\bibitem{2015userSurvey}
D.~Liu, L.~Wang, Y.~Chen, M.~Elkashlan, K.~K. Wong, R.~Schobe, and L.~Hanzo,
  ``{User Association in 5G Networks: A Survey and an Outlook},'' \emph{IEEE
  Communications Surveys \& Tutorials}, vol.~18, no.~2, pp. 1018--1044, 2016.

\bibitem{vu2017ultra}
T.~K. Vu~{et al.}, ``{Ultra-Reliable and Low Latency Communication in
  mmWave-Enabled Massive MIMO Networks},'' \emph{IEEE Communications Letters},
  vol.~21, no.~9, pp. 1--4, 2017.

\bibitem{2015optimal}
N.~Omidvar~{et al.}, ``{Optimal Hierarchical Radio Resource Management for
  HetNets with Flexible Backhaul},'' \emph{submitted to IEEE Transactions on
  Signaling Processing}, 2016.

\bibitem{2014overviewLB}
J.~Andrews, S.~Singh, Q.~Ye, X.~Lin, and H.~Dhillon, ``{An overview of load
  balancing in HetNets: Old myths and open problems},'' \emph{IEEE Wireless
  Communications}, vol.~21, no.~2, pp. 18--25, 2014.

\bibitem{rusek2013s}
F.~Rusek~{et al.}, ``{Scaling up MIMO: Opportunities and challenges with very
  large arrays},'' \emph{IEEE Signal Processing Magazine}, vol.~30, no.~1, pp.
  40--60, 2013.

\bibitem{wagner2012l}
S.~Wagner~{et al.}, ``{Large system analysis of linear precoding in correlated
  MISO broadcast channels under limited feedback},'' \emph{IEEE Transactions on
  Information Theory}, vol.~58, no.~7, pp. 4509--4537, 2012.

\bibitem{Liu2014}
A.~Liu and V.~Lau, ``{Hierarchical Interference Mitigation for Massive MIMO
  Cellular Networks},'' \emph{IEEE Transactions on Signal Processing}, vol.~62,
  no.~18, pp. 4786--4797, Sept 2014.

\bibitem{2015RZF}
Z.~Jun~{et al.}, ``Large system analysis of cognitive radio network via
  partially-projected regularized zero-forcing precoding,'' \emph{IEEE
  Transactions on Wireless Communications}, vol.~14, no.~9, pp. 4934--4947,
  2015.

\bibitem{pareto2011}
H.~Boche, S.~Naik, and M.~Schubert, ``Pareto boundary of utility sets for
  multiuser wireless systems,'' \emph{IEEE/ACM Transactions on Networking},
  vol.~19, no.~2, pp. 589--601, 2011.

\bibitem{pareto2012}
Z.~Chen~{et al.}, ``{Pareto region characterization for rate control in MIMO
  interference systems and Nash bargaining},'' \emph{IEEE Transactions on
  Automatic Control}, vol.~57, no.~12, pp. 3203--3208, 2012.

\bibitem{neely2010S}
M.~J. Neely, ``Stochastic network optimization with application to
  communication and queueing systems,'' \emph{Synthesis Lectures on
  Communication Networks}, vol.~3, no.~1, pp. 1--211, 2010.

\bibitem{beck2010seq}
A.~Beck, A.~Ben-Tal, and L.~Tetruashvili, ``A sequential parametric convex
  approximation method with applications to nonconvex truss topology design
  problems,'' \emph{Journal of Global Optimization}, vol.~47, no.~1, pp.
  29--51, 2010.

\bibitem{tran2012}
L.~N. Tran~{et al.}, ``{Fast converging algorithm for weighted sum rate
  maximization in multicell MISO downlink},'' \emph{IEEE Signal Processing
  Letters}, vol.~19, no.~12, pp. 872--875, 2012.

\bibitem{li2014e}
H.~Li, L.~Song, and M.~Debbah, ``Energy efficiency of large-scale multiple
  antenna systems with transmit antenna selection,'' \emph{IEEE Transactions on
  Communications}, vol.~62, no.~2, pp. 638--647, 2014.

\bibitem{yalmip2004}
J.~L{\"o}fberg, ``{YALMIP: A toolbox for modeling and optimization in
  MATLAB},'' in \emph{IEEE International Symposium on Computer Aided Control
  Systems Design}, New Orleans, LA, USA, 2004, pp. 284--289.

\bibitem{sdpt31999}
K.-C. Toh, M.~J. Todd, and R.~H. T{\"u}t{\"u}nc{\"u}, ``{SDPT3 - a MATLAB
  software package for semidefinite programming, version 1.3},''
  \emph{Optimization methods and software}, vol.~11, no. 1-4, pp. 545--581,
  1999.

\bibitem{mosek2015}
A.~MOSEK, ``{The MOSEK optimization toolbox for MATLAB manual, Version 7.1
  (Revision 28)},'' \emph{http://mosek. com,(accessed on March 20, 2015)},
  2015.

\bibitem{ben2001SOC}
A.~Ben-Tal and A.~Nemirovski, ``On polyhedral approximations of the
  second-order cone,'' \emph{Mathematics of Operations Research}, vol.~26,
  no.~2, pp. 193--205, 2001.

\bibitem{nguyen2015}
K.-G. Nguyen, L.-N. Tran, O.~Tervo, Q.-D. Vu, and M.~Juntti, ``{Achieving
  energy efficiency fairness in multicell MISO downlink},'' \emph{IEEE
  Communications Letters}, vol.~19, no.~8, pp. 1426--1429, 2015.

\bibitem{mo2000fair}
J.~Mo and J.~Walrand, ``Fair end-to-end window-based congestion control,''
  \emph{IEEE/ACM Transactions on Networking (ToN)}, vol.~8, no.~5, pp.
  556--567, 2000.

\bibitem{mW2014}
A.~Mustafa Riza~{et al.}, ``Millimeter wave channel modeling and cellular
  capacity evaluation,'' \emph{IEEE Journal on Selected Areas in
  Communications}, vol.~32, no.~6, pp. 1164--1179, 2014.

\bibitem{2005antenna}
C.~A. Balanis, \emph{Antenna theory: analysis and design}.\hskip 1em plus 0.5em
  minus 0.4em\relax John Wiley \& Sons, 2005.

\bibitem{Bai2014}
T.~Bai~{et al.}, ``Millimeter wave cellular channel models for system
  evaluation,'' in \emph{IEEE International Conference on Computing, Networking
  and Communications}, Honolulu, HI, USA, April 2014, pp. 178--182.

\end{thebibliography}
\begin{IEEEbiography}{Trung Kien Vu}
	received the B.Eng. degree from the School of Electronics and Telecommunications, Hanoi University of Science and Technology, Vietnam, in 2012, and the M.Sc. degree in electrical engineering from the School of Electrical Engineering, University of Ulsan, South Korea, in 2014. He is currently pursuing the D.Sc. degree with the Centre for Wireless Communications, University of Oulu, Finland. His research interests focus on heterogeneous cellular networks, massive MIMO, mm-wave communications, network planning and optimization, and ultra-reliable and low latency communications. He received the 2016 European Wireless Best Paper Award and was a recipient of the 2014 Brain Korean 21 Plus (BK21+) Scholarship.
\end{IEEEbiography}

\begin{IEEEbiography}{Mehdi Bennis}
    received his M.Sc. degree in Electrical Engineering jointly from the EPFL, Switzerland and the Eurecom Institute, France in 2002. From 2002 to 2004, he worked as a
    research engineer at IMRA-EUROPE investigating adaptive equalization algorithms for mobile digital TV. In 2004, he joined the Centre for Wireless Communications (CWC) at the University of Oulu, Finland as a research scientist. In 2008, he was a visiting researcher at the Alcatel-Lucent chair on flexible radio, SUPELEC. He obtained his Ph.D. in December 2009 on spectrum sharing for future mobile cellular systems. Currently Dr. Bennis is an Adjunct Professor at the University of Oulu and Academy of Finland research fellow. His main research interests are in radio resource management, heterogeneous networks, game theory and machine learning in 5G networks and beyond. He has co-authored one book and published more than 100 research papers in international conferences, journals and book chapters. He was the recipient of the prestigious 2015 Fred W. Ellersick Prize from the IEEE Communications Society, the 2016 Best Tutorial Prize from the IEEE Communications Society and the 2017 EURASIP Best paper Award for the Journal of Wireless Communications and Networks. Dr. Bennis serves as an editor for the IEEE Transactions on Wireless Communication.
\end{IEEEbiography}

\begin{IEEEbiography}{Sumudu Samarakoon}
	received his B. Sc. (Hons.) degree in Electronic and Telecommunication Engineering from the University of Moratuwa, Sri Lanka and the M. Eng. degree from the Asian Institute of Technology, Thailand in 2009 and 2011, respectively.
	He is currently persuading Dr. Tech degree in Communications Engineering in University of Oulu, Finland.
	Sumudu is also a member of the research staff of the Centre for Wireless Communications (CWC), Oulu, Finland.
	His main research interests are in heterogeneous networks, radio resource management, machine learning, and game theory.
\end{IEEEbiography}
\vfill

\begin{IEEEbiography}{M{\'e}rrouane Debbah}
entered the Ecole Normale Supérieure Paris-Saclay (France) in 1996 where he received his M.Sc. and Ph.D. degrees respectively. He worked for Motorola Labs (Saclay, France)  from 1999-2002 and the Vienna Research Center for Telecommunications (Vienna, Austria) until 2003. From 2003 to 2007, he joined the Mobile Communications department of the Institut Eurecom  (Sophia Antipolis, France) as an Assistant Professor. Since 2007, he is a Full Professor at CentraleSupelec (Gif-sur-Yvette, France). From 2007 to 2014, he was the director of the Alcatel-Lucent Chair on Flexible Radio. Since 2014, he is Vice-President of the Huawei France R$\&$D center and director of the Mathematical and Algorithmic Sciences Lab. His research interests lie in fundamental mathematics, algorithms, statistics, information $\&$ communication sciences research. He is an Associate Editor in Chief of the journal Random Matrix: Theory and Applications and was an associate and senior area editor for IEEE Transactions on Signal Processing respectively in 2011-2013 and 2013-2014. M{\'e}rrouane Debbah is a recipient of the ERC grant MORE (Advanced Mathematical Tools for Complex Network Engineering). He is a IEEE Fellow, a WWRF Fellow and a member of the academic senate of Paris-Saclay. He has managed 8 EU projects and more than 24 national and international projects. He received 17 best paper awards, among which the 2007 IEEE GLOBECOM best paper award, the Wi-Opt 2009 best paper award, the 2010 Newcom++ best paper award, the WUN CogCom Best Paper 2012 and 2013 Award, the 2014 WCNC best paper award, the 2015 ICC best paper award, the 2015 IEEE Communications Society Leonard G. Abraham Prize, the 2015 IEEE Communications Society Fred W. Ellersick Prize, the 2016 IEEE Communications Society Best Tutorial paper award, the 2016 European Wireless Best Paper Award and the 2017 Eurasip Best Paper Award as well as the Valuetools 2007, Valuetools 2008, CrownCom2009, Valuetools 2012 and SAM 2014 best student paper  awards. He is the recipient of the Mario Boella award in 2005, the IEEE Glavieux Prize Award in 2011 and the Qualcomm Innovation Prize Award in 2012.
\end{IEEEbiography}

\begin{IEEEbiography}{Matti Latva-aho}
	 received the M.Sc., Lic.Tech. and Dr. Tech (Hons.) degrees in Electrical Engineering from the University of Oulu, Finland in 1992, 1996 and 1998, respectively. From 1992 to 1993, he was a Research Engineer at Nokia Mobile Phones, Oulu, Finland after which he joined Centre for Wireless Communications (CWC) at the University of Oulu. Prof. Latva-aho was Director of CWC during the years $1998$-$2006$ and Head of Department for Communication Engineering until August 2014. Currently he is Professor of Digital Transmission Techniques at the University of Oulu. He serves as Academy of Finland Professor in $2017$-$2022$. His research interests are related to mobile broadband communication systems and currently his group focuses on 5G and beyond systems research. Prof. Latva-aho has published 300+ conference or journal papers in the field of wireless communications. He received Nokia Foundation Award in 2015 for his achievements in mobile communications research.
\end{IEEEbiography}
\end{document}